\documentclass[usenatbib]{mnras}
\usepackage{natbib,amssymb,amsmath,times,graphicx,url,aas_macros}
\usepackage{setspace}      % To manually set line spacing
\usepackage{epstopdf}       % to convert eps images to pdf images
\usepackage{verbatim}       % to allow multiline comments
\usepackage{bm}		  % allow shorthand to bold text
\usepackage{fleqn}		  % flush left equations
\usepackage{Wurster_macros}  % my macros
\title[The impact of non-ideal MHD on star formation]{The impact of non-ideal magnetohydrodynamic processes on discs, outflows, counter-rotation and magnetic walls during the early stages of star formation}

\author[Wurster, Bate \& Bonnell]{James Wurster$^{1,2}$\thanks{jhw5@st-andrews.ac.uk}, Matthew R. Bate$^{2}$ and Ian A. Bonnell$^{1}$\\
$^{1}$Scottish Universities Physics Alliance (SUPA), School of Physics and Astronomy, University of St. Andrews, North Haugh, St Andrews, Fife KY16 9SS, UK \\
$^{2}$School of Physics and Astronomy, University of Exeter, Stocker Rd, Exeter EX4 4QL, UK \\
}
\date{Submitted: Revised: Accepted: }
\pagerange{\pageref{firstpage}--\pageref{lastpage}} \pubyear{2021}

\begin{document}
\label{firstpage}
\bibliographystyle{mnras}
\maketitle

%255 words via BBedit; 245 words via MS Word
\begin{abstract}
Non-ideal magnetohydrodynamic (MHD) processes -- namely Ohmic resistivity, ambipolar diffusion and the Hall effect -- modify the early stages of the star formation process and the surrounding environment.  Collectively, they have been shown to promote disc formation and promote or hinder outflows.  But which non-ideal process has the greatest impact?  Using three-dimensional smoothed particle radiation non-ideal MHD simulations, we model the gravitational collapse of a rotating, magnetised cloud through the first hydrostatic core phase to shortly after the formation of the stellar core.  We investigate the impact of each process individually and collectively.  Including any non-ideal process decreases the maximum magnetic field strength by at least an order of magnitude during the first core phase compared to using ideal MHD, and promotes the formation of a magnetic wall.  When the magnetic field and rotation vectors are anti-aligned and the Hall effect is included, rotationally supported discs of $r \gtrsim 20$~au form; when only the Hall effect is included and the vectors are aligned, a counter-rotating pseudo-disc forms that is not rotationally supported.  Rotationally supported discs of $r \lesssim 4$~au form if only Ohmic resistivity or ambipolar diffusion are included.   The Hall effect suppresses first core outflows when the vectors are anti-aligned and suppresses stellar core outflows independent of alignment.  Ohmic resistivity and ambipolar diffusion each promote first core outflows and delay the launching of stellar core outflows.  Although each non-ideal process influences star formation, these results suggest that the Hall effect has the greatest influence.  
\end{abstract}

%---------------------------------------------------------------------------------------------------------------
\section{Introduction}
\label{sec:intro}

Discs and outflows are common signatures of young, low-mass stars.   For reviews, see \citet{Bachiller1996}, \citet{WilliamsCieza2011} and \citet{Bally2016}.

Recent studies and surveys have have shown that discs vary in size, dust content, morphology and even by region \citepeg{Tobin+2015,Tobin+2020,Pinte+2016,Ansdell+2016,Ansdell+2017,Ansdell+2018,Birnstiel+2018,Andersen+2019,Loomis+2020,Hendler+2020}.  Discs are also magnetised \citepeg{Rao+2014,Stephens+2014,Stephens+2017,Harris+2018}, although interpreting the magnetic field geometry is challenging since it is typically inferred from dust polarisation and magnetic fields are not the only processes to polarise dust \citepeg{Kataoka+2015,Kataoka+2017}.  Thus, discs are common, but their characteristics are heavily influenced by their environment.

Outflows are common from young stars when the accretion process is still active \citepeg{Stephens+2013,Bally2016,Bjerkeli+2016,Lee+2017jet,Maury+2018,Reipurth+2019,Busch+2020,Acretord+2020,Allen+2020,Lee2020}; some jets can have speeds up to a few 100~\kms{} \citepeg{Hartigan+2011} and extend for several thousand au.   Given that star forming regions are permeated with magnetic fields \citerev{Crutcher2012}, it is expected that magnetic fields and outflows are linked.  Indeed, outflows are magnetised, albeit the field strength is difficult to measure \citepeg{PudritzRay2019}.  Moreover, jets are expected to be magnetically launched, either as magnetocentrifugal jets \citepeg{BlandfordPayne1982} or magnetic tower jets \citepeg{Lyndenbell1996}, and are launched once enough toroidal magnetic field has accumulated near the launching region.  Although it might be expected that the outflows follow the large-scale magnetic field,  a series of observational studies suggests that the alignment between outflows and magnetic fields may be random \citerevs{HullZhang2019,PudritzRay2019}.

To reproduce these observations, numerical simulations of star formation must be performed in a magnetised medium and form discs and launch outflows.  When using ideal magnetohydrodynamics (MHD), the magnetic breaking catastrophe occurs in strong magnetic fields and discs do not form \citepeg{AllenLiShu2003,MellonLi2008,PriceBate2007}.  However, columnated outflows are launched from both the first and stellar cores \citepeg{Tomisaka2002,BanerjeePudritz2006,MachidaInutsukaMatsumoto2006,PriceTriccoBate2012,Tomida+2013,Machida2014,BateTriccoPrice2014,WursterBatePrice2018sd,MachidaBasu2019}.  These outflows are typically magnetically launched from near the (proto)star and contain a considerable toroidal component of the magnetic field; they reach speeds of a few \kms{} for first core outflows and a few tens to a few hundreds of \kms{} for stellar core outflows.  Due to computational constraints, these outflows typically reach at most a few thousand au by the end of a simulation.  

One method to form a disc in simulations with strong magnetic fields is to include non-ideal MHD (namely Ohmic resistivity, ambipolar diffusion and/or the Hall effect) to account for the low ionisation fraction in star forming regions  \citepeg{MestelSpitzer1956,NakanoUmebayashi1986,UmebayashiNakano1990}.  These processes represent how the various charged particles interact with each other, with the neutral particles and with the magnetic field.  \citet{Wardle2007} suggested that each term is dominant in a given region of the magnetic field-density phase-space.  However, \citet{Wurster2021} recently showed that all three terms (particularly ambipolar diffusion and the Hall effect) are important when modelling protostellar discs, and that defining regions of the phase-space where each process is dominant was an over-simplification (see also references therein).   Ohmic resistivity and ambipolar diffusion are both diffusive terms that act to locally weaken the magnetic field, whereas the Hall effect is a dispersive term which changes the direction of the magnetic field \citep[i.e. a `vector evolution'; e.g.][]{Wardle2004}.  Given the vector evolution of the magnetic field, the Hall effect is sensitive to the polarity of the magnetic field, particularly if the gas is already rotating \citepeg{BraidingWardle2012acc}.

Under some conditions, ambipolar diffusion alone is enough to promote disc formation \citepeg{Masson+2016,Hennebelle+2016,Vaytet+2018,Marchand+2020}\footnote{These discs are 10s of au; other studies have have found only small discs of a few au form when using ambipolar diffusion \citepeg{TomidaOkuzumiMachida2015,Tsukamoto+2015oa}.}, however, the Hall effect is typically required to consistently form a protostellar disc \citepeg{Tsukamoto+2015hall,Tsukamoto+2017,\wpb2016,\wbp2018hd}.  As we previously showed in \citet{\wbp2018hd}, when the Hall effect is included and a large disc forms, outflows are suppressed.  With the exception of \citet{\wpb2016}, our previous studies included Ohmic resistivity, ambipolar diffusion and the Hall effect since all three are important for star formation, so we were unable to determine which process(es) were responsible for suppression of the outflows.  The numerical studies that focused on the outflows either modelled ideal MHD \citepeg{PriceTriccoBate2012,BateTriccoPrice2014} or included Ohmic resistivity only \citepeg{Machida2014,MachidaBasu2019}.  Thus, one of the goals of this current study is to determine what process is responsible for suppressing outflows.  

The strengths of the non-ideal MHD processes are dependent on microphysics, such as cosmic ray ionisation rate, the included chemical species and the grain size/grain size distribution \citepeg{PadovaniGalliGlassgold2009,Padovani+2014,Zhao+2016,ZhaoCaselliLi2018,Zhao+2020,Zhao+2021,Bai2017,\wbp2018sd,\wbp2018ion,Xu+2019,Tsukamoto+2020}.  The dust grain distribution has been an area of recent focus, and it was found that even when including the non-ideal MHD processes, disc formation was suppressed in the presence of very small grains \citepeg{LiKrasnopolskyShang2011,Zhao+2020}.  When using an MRN grain size distribution \citep*{MathisRumplNordsieck1977}, a minimum grain size of $a_\text{g,min} = 0.03$~$\mu$m led to a Hall-dominated regime \citepeg{ZhaoCaselliLi2018,Koga+2019} while a larger minimum grain size of $a_\text{g,min} = 0.1$~$\mu$m led to an ambipolar diffusion-dominated regime \citepeg{Zhao+2016,Dzyurkevich+2017}.  \citet{Zhao+2021} found that in each regime (Hall- or ambipolar-dominated), the non-dominant term had a minimal influence on the resulting disc formation.  Therefore, the grain size affects the importance of each non-ideal process and ultimately the evolution of the environment during the star formation process.  Grain size distribution is thus added to a long list of environmental conditions (e.g. magnetic field strength, magnetic field alignment, rotation rate, turbulence, thermal support, etc...) that affect star formation.  

Although discs and outflows are the most obvious signatures of star formation, under certain circumstances, counter-rotating regions and magnetic walls also form.  The Hall effect induces a rotational velocity  \citepeg{KrasnopolskyLiShang2011,LiKrasnopolskyShang2011,BraidingWardle2012acc,Tsukamoto+2015hall,Tsukamoto+2017,\wpb2016,Zhao+2020}, where the direction of rotation is dependent on the magnetic field's polarity.  In cores that are already rotating, this either spins up or spins down the gas; in many cases, this leads to well-defined counter-rotating regions \citepeg{Tsukamoto+2015hall,\wpb2016,Zhao+2020,Zhao+2021}, where the characteristics and longevity of these counter-rotating regions are dependent on the initial environment.  Unfortunately, these regions will be difficult to observe since their signal is expected to be weak and their line-of-sight velocity component may be confused with infall motions \citep{Yen+2017arcs}.

Finally, charged and neutral particles behave differently in a magnetic field.  During the collapse, the magnetic field decelerates the charged particles but not the neutral gas.  This leads to a less severe pinching of the magnetic field near the core compared to ideal MHD \citepeg{WursterLi2018}.  However, the deceleration of the charged particles is not uniform, with particles closer to the centre (where the magnetic field is stronger) undergoing a greater deceleration. This leads to a pile-up of the magnetic flux outside of the core, leading to the formation of a so-called `magnetic wall' \citepeg{TassisMouschovias2005b,TassisMouschovias2007a,TassisMouschovias2007b}.  In their study, \citet{TassisMouschovias2005b} found a rapid steepening of the magnetic field strength that increased by a few orders of magnitude over a very short distance as a result of ambipolar diffusion differentially decelerating the electrons.  Similarly, \citet{\wbp2018ff} found that the magnetic flux built up in a torus around the centre of the core when the three non-ideal MHD terms were included.  Interior to the torus, however, was a spiral structure, suggesting that non-ideal MHD lead to more than just a simple pile-up of magnetic flux in a simple wall or torus.

In this study, we investigate the effect that each non-ideal process has on star formation, with a focus on the environmental impacts including discs, outflows, counter-rotation and magnetic walls.  Using 3D smoothed particle radiation magnetohydrodynamic simulations, we model the gravitational collapse of a molecular cloud core through the first and stellar core phases in a magnetised medium.  Four simulations each include only one non-ideal MHD process so that its effect can be disentangled from our previous simulations that included the three effects.   The concept of this study is similar to \citet{WursterPriceBate2016}, however, this study uses radiation hydrodynamics rather than a barotropic equation of state\footnote{See \citet{Bate2011} for a discussion and comparison of radiation hydrodynamics versus a barotropic equation of state in hydrodynamic star formation simulations; \citet{LewisBate2018} perform a similar comparison using ideal MHD simulations.  See additional discussions in \citet{Tomida+2013,TomidaOkuzumiMachida2015}.} and excludes sink particles\footnote{See \citet{MachidaInutsukaMatsumoto2014} for a discussion on how sink size affects the star forming environment.} so that we can model the formation of the stellar core and the launching of the stellar core outflow.  In \secref{sec:methods}, we present our methods, and in \secref{sec:ics}, we present our initial conditions.  In \secref{sec:results}, we present and discuss our results, and in \secref{sec:disc} we discuss our results in a broader context.  We conclude in \secref{sec:conc}. 

%----------------------------------------------------------------------------------------------------------------
\section{Methods}
\label{sec:methods}

Our method is identical to that from our previous studies \citepeg{WursterBatePrice2018sd,WursterBatePrice2018hd,WursterBatePrice2018ff}.  We solve the self-gravitating, radiation non-ideal magnetohydrodynamics equations using \textsc{sphNG}, which is a three-dimensional Lagrangian smoothed particle hydrodynamics (SPH) code that originated from \citet{Benz1990}.   This code has been substantially modified to include (e.g.) a consistent treatment of variable smoothing lengths \citep{PriceMonaghan2007}, individual time-stepping \citep{BateBonnellPrice1995}, radiation as flux limited diffusion \citep{WhitehouseBateMonaghan2005,WhitehouseBate2006}, magnetic fields \citep[for a review, see][]{Price2012}, and non-ideal MHD \citep{WursterPriceAyliffe2014,WursterPriceBate2016}.  For stability of the magnetic field, we use the source-term subtraction approach \citep{BorveOmangTrulsen2001}, constrained hyperbolic/parabolic divergence cleaning \citep{TriccoPrice2012,TriccoPriceBate2016}, and artificial resistivity \citep[as described in][]{Price+2018phantom}.  For a more detailed description, see \citet{WursterBatePrice2018sd}\footnote{Note that \citet{WursterBatePrice2018sd} used the artificial resistivity from \citet{TriccoPrice2013} rather than \citet{Price+2018phantom}.}.

The non-ideal MHD coefficients are self-consistently calculatedly using the \textsc{Nicil} library \citep{Wurster2016}\footnote{Models ohaMHD$\pm$ from our previous studies use \textsc{Nicil} v1.2.1 and the remaining models use \textsc{Nicil} v1.2.5; performance and stability was enhanced between the versions, with only trivial effects on the resulting resistivities.}.  At low temperatures, light and heavy metals and dust grains can be ionised by cosmic rays.  There are three grain populations of radius $a_\text{g} = 0.1$~$\mu$m that differ only in charge ($Z = 0, \ \pm1$); the analysis in \citet{Wurster2021} suggest that this grain size corresponds to neither the Hall- or ambipolar diffusion-dominate regime as discussed in \secref{sec:intro}.  At high temperatures ($T \gtrsim 1000$~K), the gas can become thermally ionised.  We selectively include the three non-ideal MHD terms that are important for star formation: Ohmic resistivity, ambipolar diffusion and the Hall effect.  Ohmic resistivity is calculated implicitly (see the appendix of \citealt{WursterBatePrice2018sd}), while the Hall effect and ambipolar diffusion are calculated explicitly.  We use a default non-ideal time-step coefficient of  $C_\text{nimhd} = 1/2\pi$, however for numerical stability, the coefficient is decreased to $C_\text{nimhd} = 1/4\pi$ for the two models that include only the Hall effect.

%----------------------------------------------------------------------------------------------------------------
\section{Initial conditions}
\label{sec:ics}

Our initial conditions are identical to those used in our previous studies \citepeg{BateTriccoPrice2014,WursterBatePrice2018sd,WursterBatePrice2018hd,WursterBatePrice2018ff}.  We initialise a spherical core of mass 1~M$_{\odot}$, radius $R_\text{c} = 4\times10^{16}$~cm, uniform density $\rho_0 = 7.42\times 10^{-18}$~\gpercc{} and initial (isothermal) sound speed $c_\text{s} = \sqrt{p/\rho}= 2.2\times 10^{4}$~cm~s$^{-1}$; this corresponds to a ratio of thermal-to-gravitational potential energy of $\alpha_0 = 0.36$.  We include solid body rotation about the $z$-axis of $\Omega_0 = 1.77 \times 10^{-13}$~rad s$^{-1}$ (i.e. $\bm{\Omega_0} = \Omega_0 \hat{\bm{z}}$), which corresponds to a ratio of rotational-to-gravitational potential energy of $\beta_\text{r} = 0.005$.  The core is placed in pressure equilibrium with a warm, low-density ambient medium of edge length $4R_\text{c}$ and density contrast of 30:1; magnetohydrodynamic forces are periodic across the boundary of this box but gravitational forces are not.

The entire system is threaded with a uniform vertical magnetic field of strength $B_0 = 163\mu$G, which is equivalent to a mass-to-flux ratio of $\mu_0 = 5$ in units of the critical mass-to-flux ratio \citepeg{Mestel1999,MaclowKlessen2004}.  For the non-ideal MHD processes, we use the canonical cosmic ray ionisation rate of \zetaeq{-17}  \citep{SpitzerTomasko1968}.

Our models are listed in \tabref{table:models}, along with the included non-ideal processes and the initial magnetic field orientation since the Hall effect is sensitive to the sign of the magnetic field vector.  Models iMHD and ohaMHD$\pm$ were analysed and published in previous studies, as listed in the fourth column.

\begin{table}
\centering
\begin{tabular}{l l l l}
\hline
Name        & Non-ideal effect(s)            & $B$-field                 & Source \\
\hline
iMHD        & none                                  & $-B_0\hat{\bm{z}}$ & \citet{WursterBatePrice2018hd,WursterBatePrice2018ff} \\
oMHD       & Ohmic resistivity                              & $-B_0\hat{\bm{z}}$ & this study \\ 
hMHD+     & Hall effect                                  & $+B_0\hat{\bm{z}}$ & this study \\
hMHD-      & Hall effect                         & $-B_0\hat{\bm{z}}$ & this study \\
aMHD       & ambipolar diffusion                          & $-B_0\hat{\bm{z}}$ & this study \\
ohaMHD+ & Ohmic, Hall, ambipolar & $+B_0\hat{\bm{z}}$ & \citet{WursterBatePrice2018hd,WursterBatePrice2018ff} \\
ohaMHD-  & Ohmic, Hall, ambipolar & $-B_0\hat{\bm{z}}$  & \citet{WursterBatePrice2018hd} \\ 
\hline
\end{tabular}
\caption{A summary of our models.  We cite the source and previous literature for the models that were not performed explicitly for this study.  Only the Hall effect is sensitive to the sign of the magnetic field vector.}
\label{table:models}
\end{table}

Each model includes $3\times10^6$ equal mass SPH particles in the spherical core and an additional $1.5 \times 10^6$ particles in the warm medium.  At this resolution, it takes $6\times10^4$ -- $3\times10^5$ cpu-h for a simulation to run, with iMHD taking the shortest amount of time and ohaMHD- taking the longest.

In our previous studies, we refer to \textit{non-ideal} models as those including Ohmic resistivity, ambipolar diffusion and the Hall effect (e.g. ohaMHD$\pm$); we continue to use this term here as well.   We collectively refer to Ohmic resistivity and ambipolar diffusion as the diffusive terms.

%----------------------------------------------------------------------------------------------------------------
\section{Results}
\label{sec:results}

We follow the gravitational collapse of the cloud core through the first hydrostatic core phase (\rhoxrange{-12}{-8}) and through the second collapse phase (\rhoxrange{-8}{-4}) to the formation of the stellar core.  We define the stellar core formation to occur at density \rhoxeq{-4}, and we define this time to be \dtsczero.  We continue to follow the evolution for another 8~mo for iMHD and for another 4~yr (i.e. until \dtsc{4}) for the remainder of the models.   The short evolution time after the formation of the stellar core is due to computational limitations, where the evolution since stellar core formation requires similar or more computational resources than the evolution to stellar core formation.  This short evolution time of the stellar core means that we cannot determine the lifespan of the features that form near the formation of the stellar core, nor can we comment on whether or not features may develop later.  

%----
\subsection{General evolution}
\begin{figure} 
\centering
\includegraphics[width=\columnwidth]{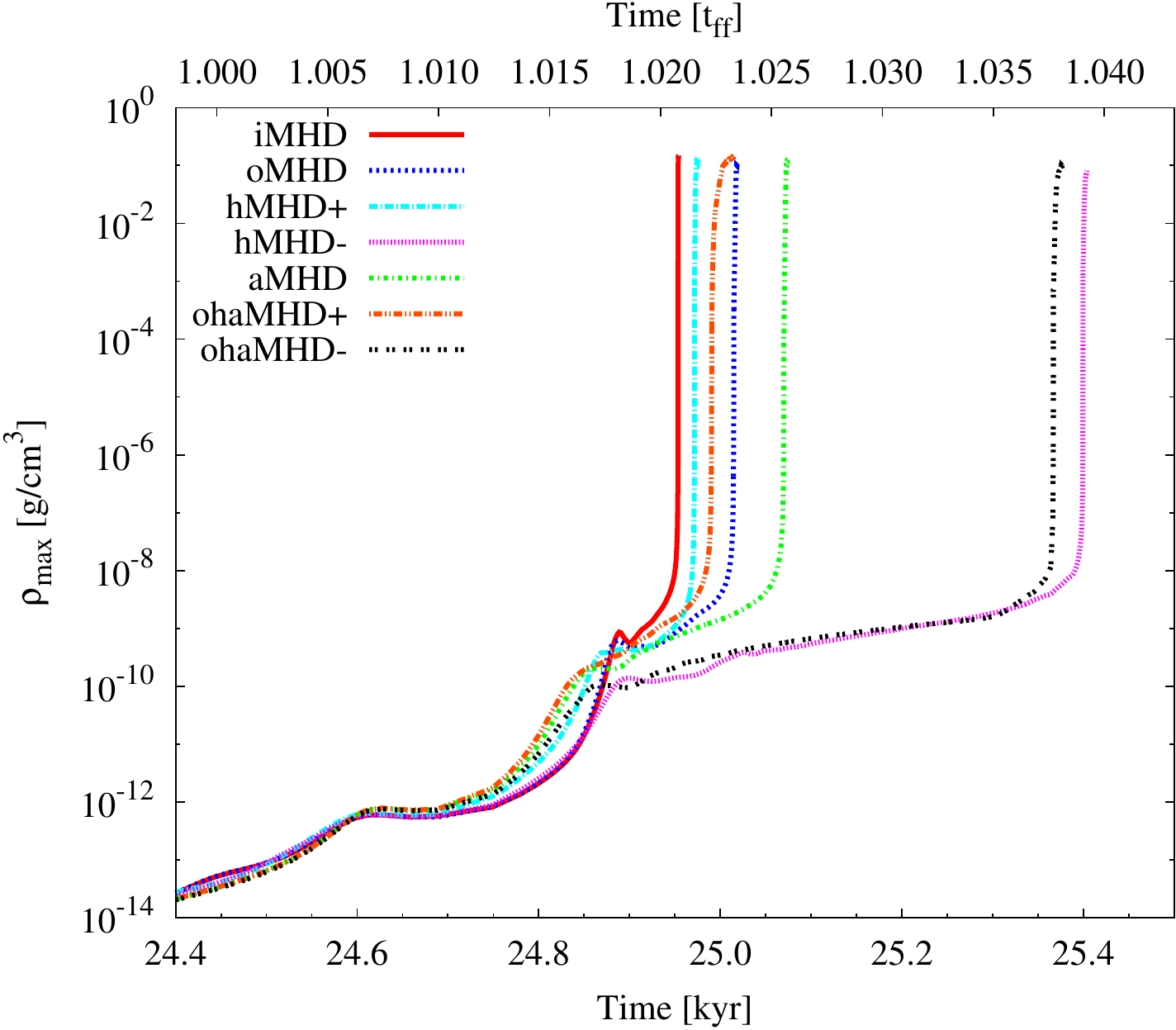} %made on Dial
\caption{Evolution of the maximum density with time.  Including non-ideal MHD processes delays the collapse to stellar densities relative to the ideal MHD model.  Models hMHD$\pm$ produce the longest and shortest increases in collapse time over iMHD.  Including the diffusive processes (Ohmic resistivity and ambipolar diffusion) tempers the extremes of the Hall-only models.}
\label{fig:rhoVtime}
\end{figure}

As a cloud core collapses to form a star, there are many processes that oppose gravity to delay the collapse, including magnetic fields.  The strength of the magnetic fields \citepeg{BateTriccoPrice2014,MachidaHiguchiOkuzumi2018}, the strength of the non-ideal MHD processes \citepeg{\wbp2018sd,\wbp2018ion,\wbp2018hd,\wbp2018ff} and even which non-ideal processes are included \citepeg{Tomida+2013,TomidaOkuzumiMachida2015,Tsukamoto+2015oa,Tsukamoto+2015hall,\wpb2016} all affect the collapse.  \figref{fig:rhoVtime} shows the time evolution of the maximum density of each model starting near the beginning of the first core phase. The lifespan of the first core varies between \sm200 -- 600~yr, depending on the model; each model spends an additional \sm200~yr in the density range \rhoxrange{-13}{-12} just prior to entering the first core phase.  Model iMHD has the shortest first core lifespan, thus adding any non-ideal processes increases its lifespan.  By including the Hall effect with the anti-aligned vectors (i.e. hMHD- and ohaMHD-), the lifespan increases to $\gtrsim 600$~yr. 

These first core lifespans are shorter than the typically quoted lifespans of \sm1000~yr, although the lifespan can range from a few 100 to a few 1000 yr. The lifetimes of first cores strongly depend on the angular momentum of the molecular cloud core \citepeg{Bate2011}.  Although our non-ideal MHD calculations all begin with the same angular momentum, due to varying amounts of magnetic angular momentum transport, the inner parts of the cloud evolve to contain differing amounts of angular momentum which similarly affects the first core lifespans (see \figsref{fig:rhoVtime}{fig:LVrho}).  In addition to variations in magnetic field evolution due to non-ideal effects, increased first core lifespans can be caused by modelling weaker magnetic fields \citepeg{Commercon+2012,BateTriccoPrice2014}, including turbulence \citepeg{LewisBate2018}, including more initial mass \citepeg{Tomida+2010llc}, or starting with a centrally condensed gas cloud \citepeg{Tomida+2013,TomidaOkuzumiMachida2015,WursterBate2019}.  In general, any process that decreases the accretion rate onto the first core increases its lifespan \citepeg{SaigoTomisaka2006,SaigoTomisakaMatsumoto2008,Tomida+2010llc,Matsushita+2017}.  Therefore, the lifespan of the first core is heavily dependent on its environment.  Although longer first core lifespans would permit more time for the magnetic field to evolve, its evolution would be degenerate with the processes that caused the increased lifespan.
%intro of \citet{Tsitali+2013} is/was useful

\begin{figure} 
\centering
\includegraphics[width=\columnwidth]{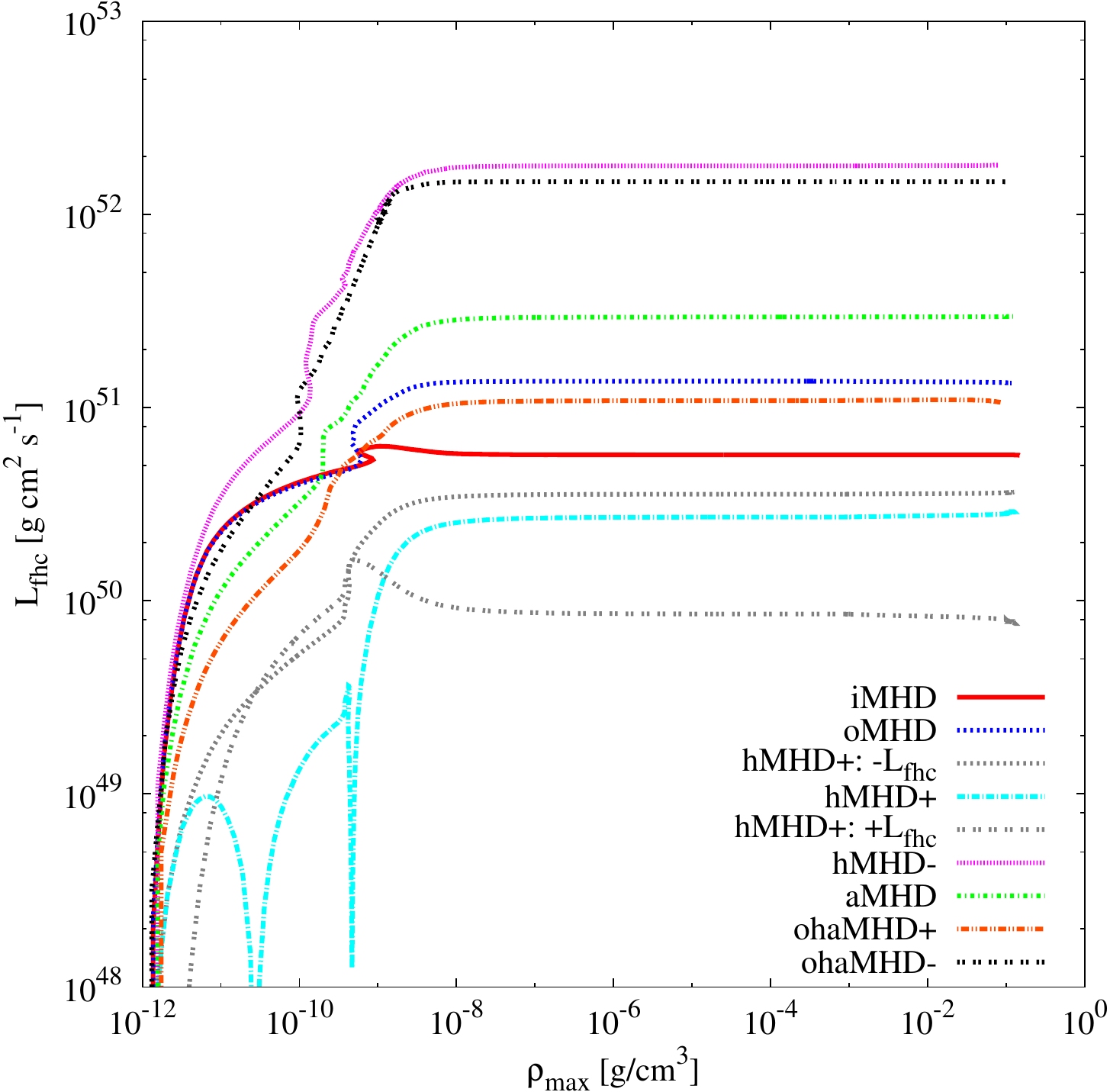} %made on Dial
\caption{Evolution of the angular momentum in the first hydrostatic core (defined as the gas with \rhoge{-12}), against maximum gas density, which is used as a proxy for time.  With the exception of hMHD+, the amount of angular momentum in the first hydrostatic core is directly related the time it takes to collapse to stellar densities, where the cores that have greater angular momentum reside longer in the first core phase.  Counter-rotating regions have formed in hMHD+ (further discussed in \secref{sec:crotation}); the three bottom curves represent hMHD+, where the angular momentum was calculated (from top to bottom) using the counter-rotating gas in the first core, all the gas, and the prograde gas, respectively.  Although there is less net angular momentum in the first core of hMHD+ than iMHD, it resides in the first core phase slightly longer. }
\label{fig:LVrho}
\end{figure}

There are already slight differences in the time the models enter the first core phase, with the evolution slowly starting to diverge around 20~kyr from the beginning of the collapse.  As has been previously shown \citepeg{\wbp2018sd,\wbp2018hd,\wbp2018ff}, it is during the first core phase that models undergo significant divergence, despite this being a short phase.  These differences in collapse time can be explained by the amount of angular momentum in the first hydrostatic core, as shown in \figref{fig:LVrho}; in this figure and several remaining figures in the text, we use maximum density as a proxy for time which yields better comparisons amongst the models.  

% Approx time in the FHC (1e-12 to 1e-8):
%iMHD :  24.95 - 24.76 = 190yr   1
%oMHD : 25.01 - 24.76 = 250yr   3
%aMHD : 25.06 - 24.71 = 350yr   5
%hMHD+ : 24.97 - 24.73 = 240yr  2
%hMHD- : 25.39 - 24.75 = 640yr   7
%ohaMHD+ : 24.99 - 24.71 = 280yr (enters first, only slightly before aMHD)  4
%ohaMHD- : 25.35 - 24.72 = 630yr  6

The differences in angular momentum are a direct result of how the varying non-ideal processes affect the magnetic field.  Ohmic resistivity and ambipolar diffusion are both diffusive terms that weaken the magnetic field, permitting more angular momentum to remain in the first core \citep[as previously shown in][]{TomidaOkuzumiMachida2015,Tsukamoto+2015oa}.  This delays the collapse compared to the ideal collapse.  The divergence from iMHD naturally occurs earlier for aMHD than oMHD since the collapse must first pass through the regime where the respective process is effective.   Since ambipolar diffusion is more effective at lower densities than Ohmic resistivity, aMHD has a greater divergence from iMHD, both in terms of collapse time and angular momentum in the first core. 

The Hall effect's vector evolution of the magnetic field strongly influences angular momentum budget of the first core by promoting or hindering the transport of angular momentum, depending on the initial orientation of the magnetic field \citepeg{Wardle2004,BraidingWardle2012acc}.  When the rotation and magnetic field vectors are anti-aligned, the Hall effect effectively contributes significantly to the angular momentum; hence, hMHD- takes the longest time to collapse and has the greatest amount of angular momentum in the first core.  

When the rotation and magnetic field vectors are aligned, the Hall effect extracts angular momentum from the central regions and counter-rotating regions form (see \secref{sec:crotation}); the bottom three curves in \figref{fig:LVrho} show the angular momentum in the first core of hMHD+ calculated using the counter-rotating gas, all the gas, and the prograde gas.  It is clear that the bulk of the angular momentum in the first core in this model is from the counter-rotating gas.  Despite hMHD+ collapsing slightly slower than iMHD (\figref{fig:rhoVtime}), neither the net angular momentum nor the angular momentum of the counter-rotating gas is greater than the angular momentum of the first core of iMHD.  Therefore, the delay in the collapse of this model is a result of the gas changing orbital direction rather than simply increasing its azimuthal velocity as in the other models.

Including all three non-ideal effects reduces the extremes obtained by hMHD$\pm$.  Since Ohmic resistivity and ambipolar diffusion diffuse the magnetic field, there is less that can be dispersed by the Hall effect.  This results in a slightly faster collapse and slightly less angular momentum in the first hydrostatic core for ohaMHD- compared to hMHD-.  In ohaMHD+, the Hall effect extracts angular momentum but the diffusive terms reduce the magnitude of the extraction.  In this model, the Hall effect does not extract so much angular momentum as to generate counter-rotating regions (see \secref{sec:crotation}), nor does it extract enough for this core to have less angular momentum than iMHD.  Indeed, the final angular momentum in the core of ohaMHD+ is approximately twice the angular momentum in iMHD.

\begin{figure} 
\centering
\includegraphics[width=\columnwidth]{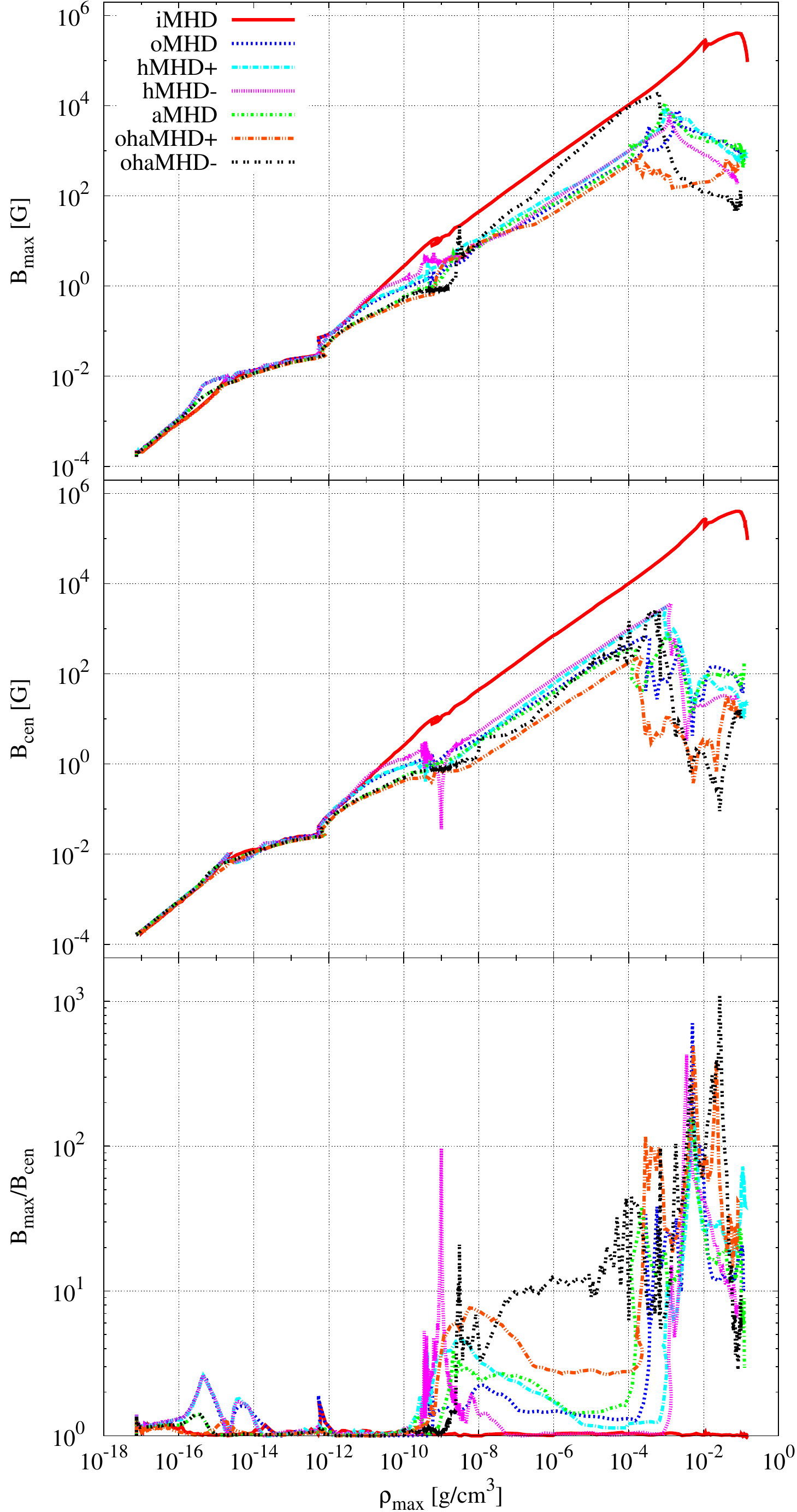} %made on Dial
\caption{Evolution of the maximum (top) and central (middle) magnetic field strengths and the ratio between them (bottom).  Including non-ideal effects decreases the maximum and central magnetic field strengths relative to the ideal MHD model, and decouples the maximum magnetic field strength from the central field strength through the formation of a magnetic wall (see \secref{sec:wall}), which occurs during the first core phase.}
\label{fig:BVrho}
\end{figure}

The top panel of \figref{fig:BVrho} shows the evolution of the maximum magnetic field strength.  As previously shown \citepeg{\wbp2018sd,Vaytet+2018}, adding non-ideal MHD effects decreases the maximum magnetic field strength, starting during the first core phase.  Adding any non-ideal process decreases the maximum magnetic field strength by at least an order of magnitude compared to iMHD by the formation of the stellar core at \rhoxeq{-4}; the exception is ohaMHD- which has already formed an $m=2$ instability followed by a protostellar disc (see \secref{sec:discs}).  The growth rate of the magnetic field strength during the second collapse phase is similar for all models (except ohaMHD-); during the second collapse phase, hMHD$\pm$ have similar maximum field strengths, which is only slightly greater than the strengths of oMHD and aMHD.

The middle panel of \figref{fig:BVrho} shows the evolution of the central magnetic field strength (coincident with the maximum gas density), and the bottom panel shows the ratio between the maximum and central strengths.  During the first core phase, the maximum strength decouples from the centre for all models that include non-ideal processes, indicating the formation of a magnetic wall.  The decoupling and the magnetic wall are further discussed in \secref{sec:wall}.

%---
\subsection{Discs}
\label{sec:discs}
Using these initial conditions, we have previously shown that iMHD never forms a disc, ohaMHD+ forms a \sm1.5~au disc after the end of the first core phase, and ohaMHD- forms a \sm~25~au disc during the first core phase \citep{\wbp2018hd}.  However, which of the non-ideal MHD processes is responsible for these discs?  In \citet{\wpb2016}, we concluded that the Hall effect is responsible for disc formation, while other studies suggest only ambipolar diffusion is required \citepeg{TomidaOkuzumiMachida2015,Tsukamoto+2015oa,Masson+2016,Hennebelle+2016,Vaytet+2018}.  

\begin{figure*} 
\centering
\includegraphics[width=0.5895\textwidth]{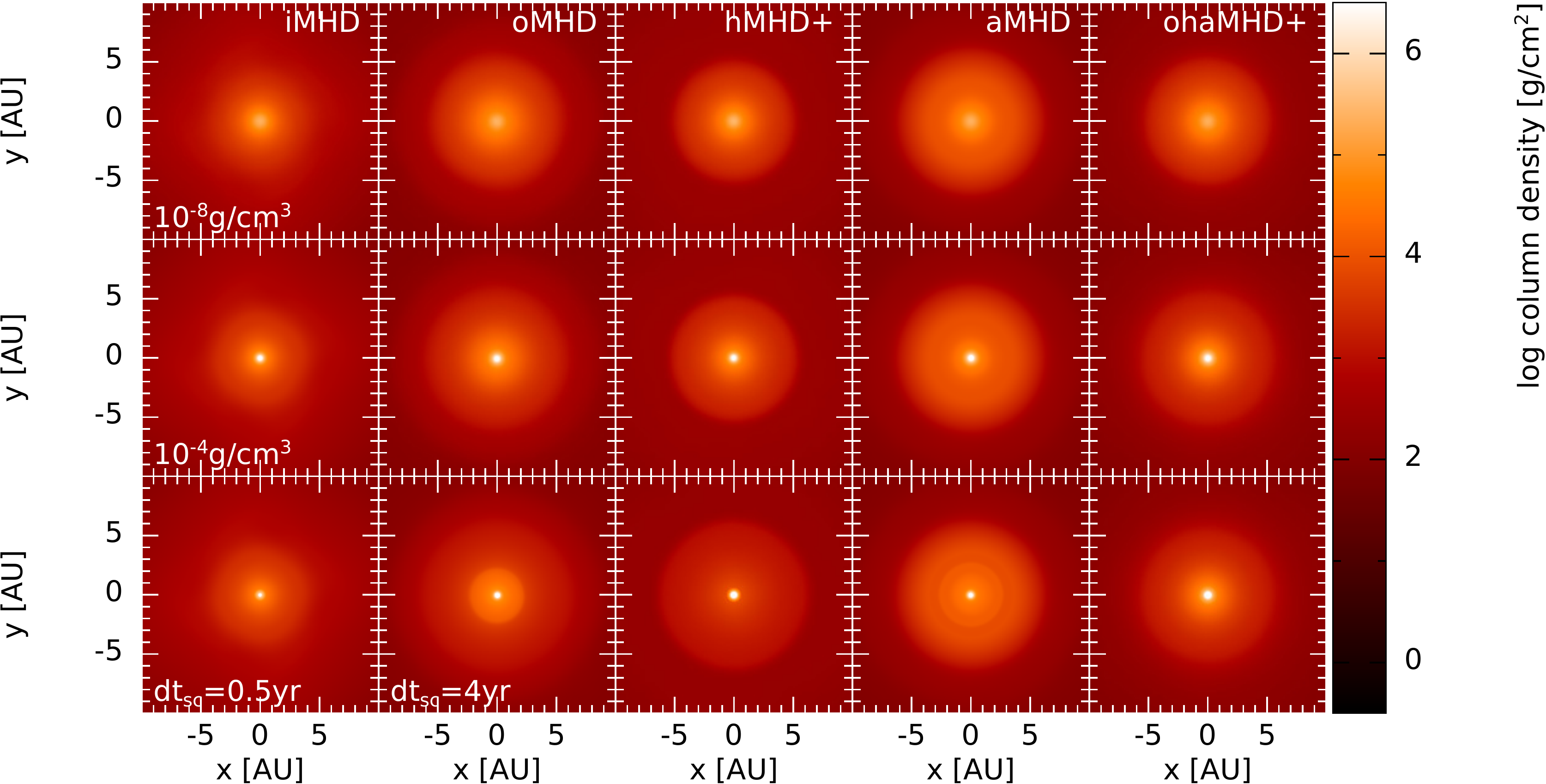}
\includegraphics[width=0.306\textwidth]{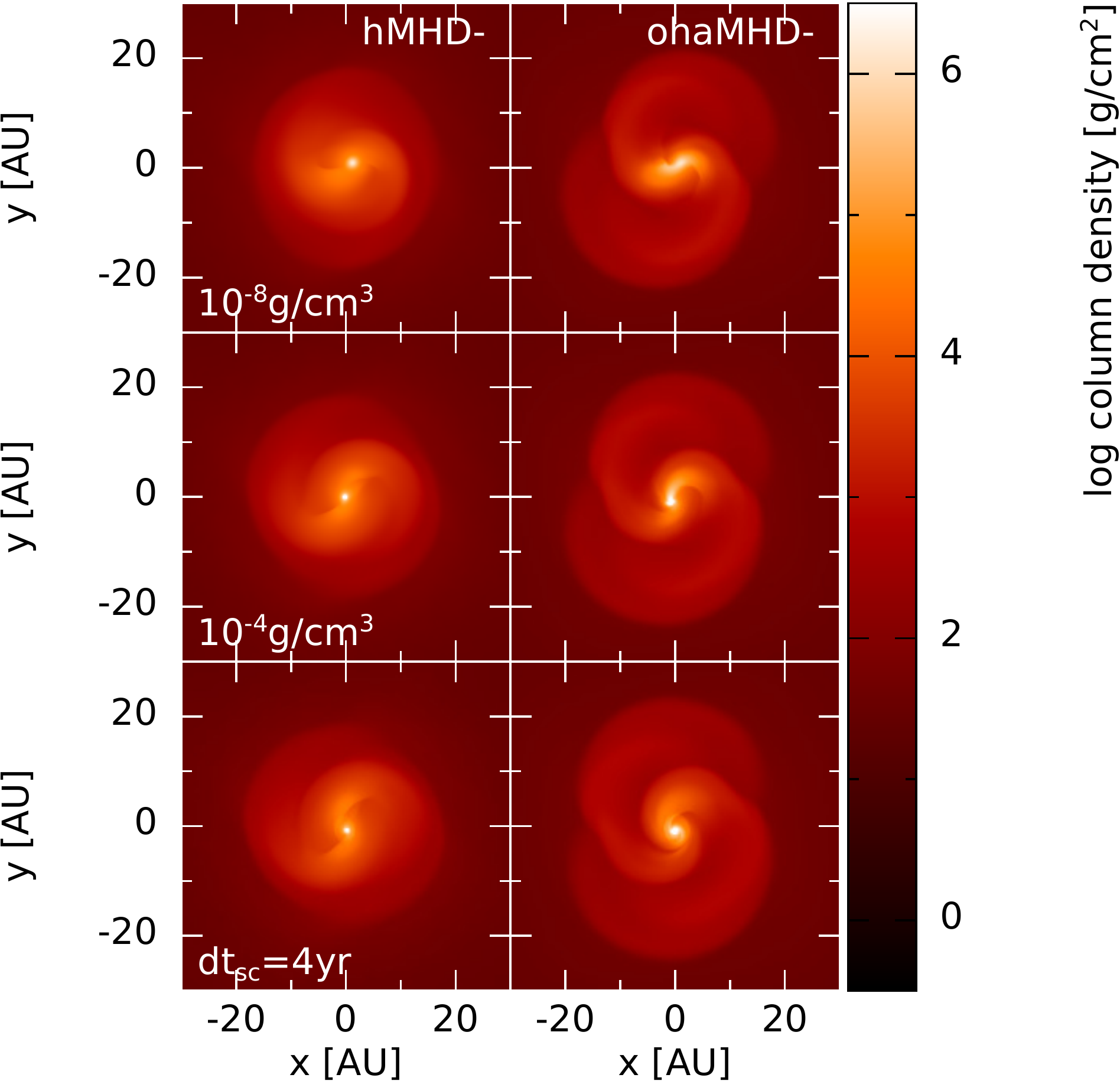}
\caption{Gas column density near the end of the first hydrostatic core phase (top row), at the formation of the stellar core (middle row) and 4~yr after the formation of the stellar core (0.5~yr for iMHD; bottom row).  The left and right panels have different spatial scales.  Large \sm20~au discs form in the first core phase when the Hall effect is included and the magnetic field and rotation vectors are initially anti-aligned (right-hand panel). }
\label{fig:disc}
\end{figure*}

\figref{fig:disc} shows the gas column density at three times during the evolution.  It is immediately clear that the Hall effect with the anti-aligned magnetic field must be included for an obvious disc to form (i.e. hMHD- and ohaMHD-; right-hand panel).  These discs form during the first core phase and have $r \gtrsim 20$~au.  The angular momentum in the first core of these two models is \sm2 times larger than aMHD, which is the model with the next largest angular momentum (\figref{fig:LVrho}).  The spiral structure in these discs forms from a classical rotationally-unstable rapidly-rotating polytrope type instability (e.g. \citealt{Tohline1980} which uses an isothermal equation of state, \citealt{Bonnell1994} and \citealt{BonnellBate1994} which use a polytropic equation of state,  \citealt{Bate1998} which uses a barotropic equation of state, or \citealt{Bate2011} which uses radiative transfer) and does not depend on the presence of a magnetic field.  Just prior to the formation of the spiral structure, the ratio of rotational-to-gravitational energy of the first core exceeds the critical value required for dynamic instability to nonaxisymmetric perturbations \citep[i.e. $\beta_\text{r} > 0.274$;][]{Durisen+1986}.  In the models that do not form large, unstable discs (i.e. the models shown in the left-hand panel of \figref{fig:disc}), the ratio is well below this critical value.

\begin{figure} 
\centering
\includegraphics[width=\columnwidth]{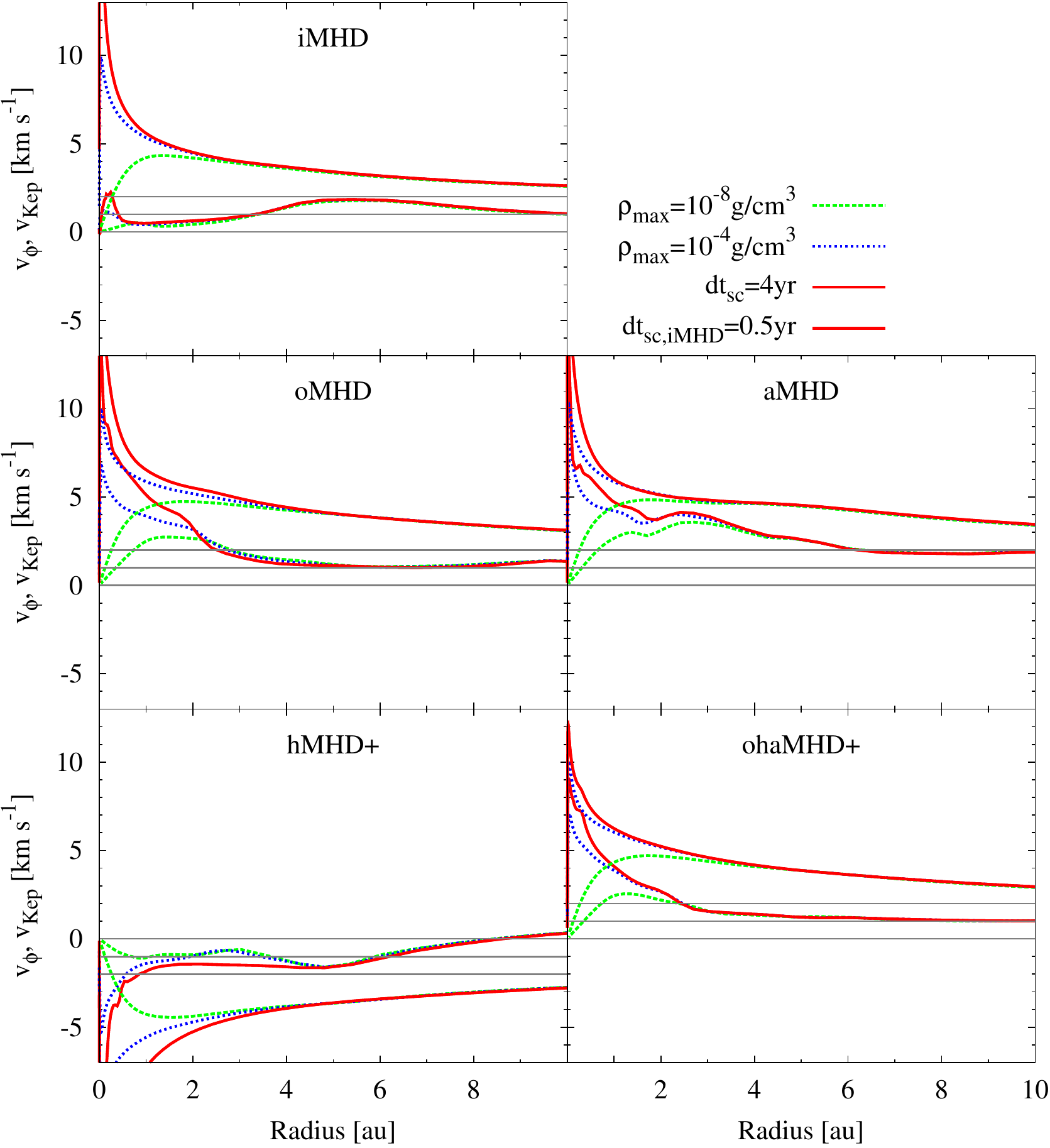}
\includegraphics[width=\columnwidth]{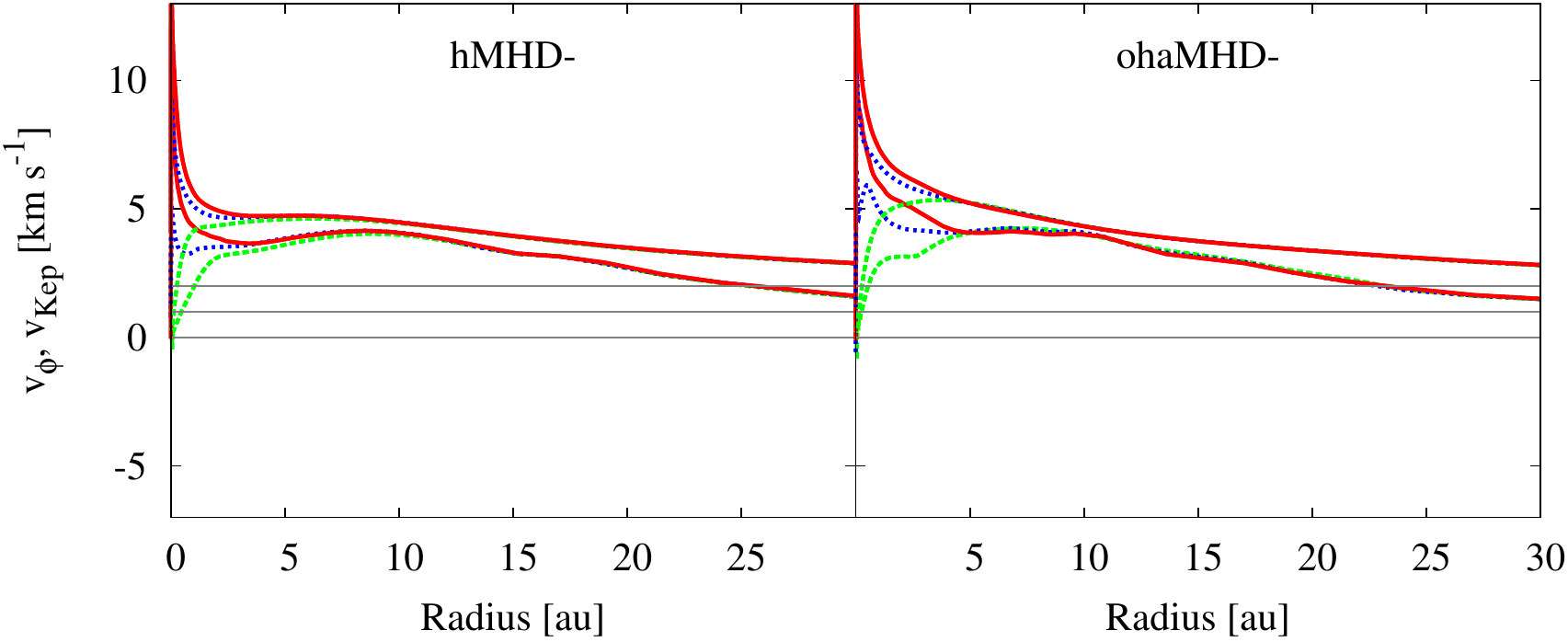}
\caption{The azimuthally averaged Keplerian and azimuthal velocities of the gas within $20^\circ$ of the mid-plane at the same epochs as in \figref{fig:disc}.  For all panels except hMHD+, the smooth, upper lines are the Keplerian velocity, $v_\text{Kep}$, the lower lines are the azimuthal velocity $v_\phi$, and the grey references lines are $v = 0, 1, 2$~\kms; for hMHD+, the smooth, lower lines are $v_\text{Kep}$, the upper lines are $v_\phi$, and the grey references lines are $v = 0, -1, -2$~\kms.  All mid-planes are rotating at sub-Keplerian velocities.  Models hMHD- and ohaMHD- are rotating with greater than half the Keplerian velocity providing evidence for their discs, while iMHD and hMHD- are rotating much slower than $v_\text{kep}$, suggesting the absence of a disc.} 
\label{fig:disc:v}
\end{figure}

\figref{fig:disc:v} shows the azimuthally averaged Keplerian and azimuthal velocities in the mid-plane.  All mid-planes are rotating at sub-Keplerian velocities, with rotational speeds of a few \kms{}.  Models hMHD- and ohaMHD- are rotating the fastest, with $v_\phi > 0.5v_\text{Kep}$, providing further evidence of their disc.  Notably, hMHD+ is rotating in the opposite direction to the initial rotation of the cloud and to remaining models (see \secref{sec:crotation}), with a slow rotation of $|v_\phi| < 2$~\kms{}.  
These rotation profiles suggest that discs have not formed in iMHD or hMHD+.

\begin{figure} 
\centering
\includegraphics[width=\columnwidth]{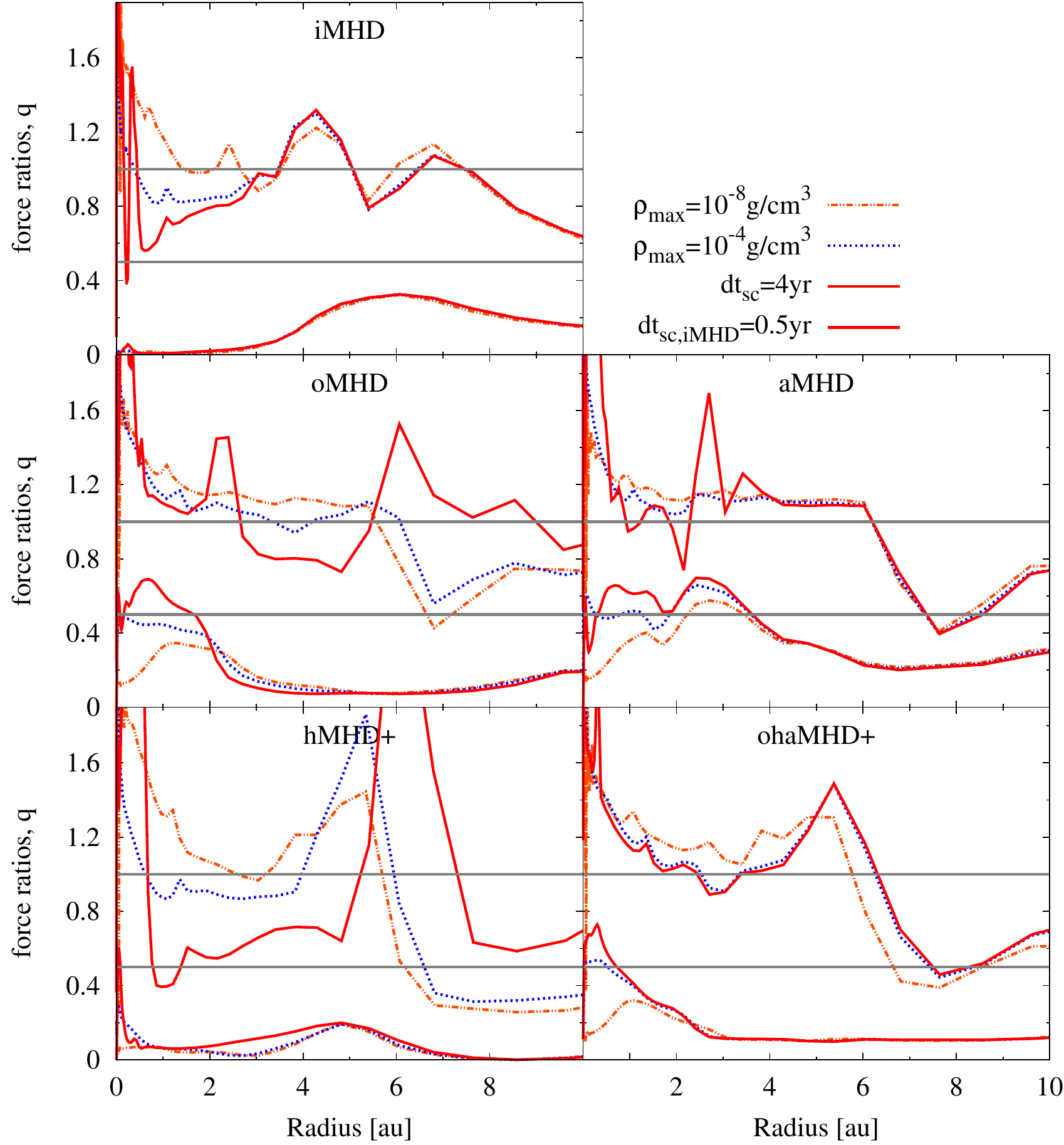}
\includegraphics[width=\columnwidth]{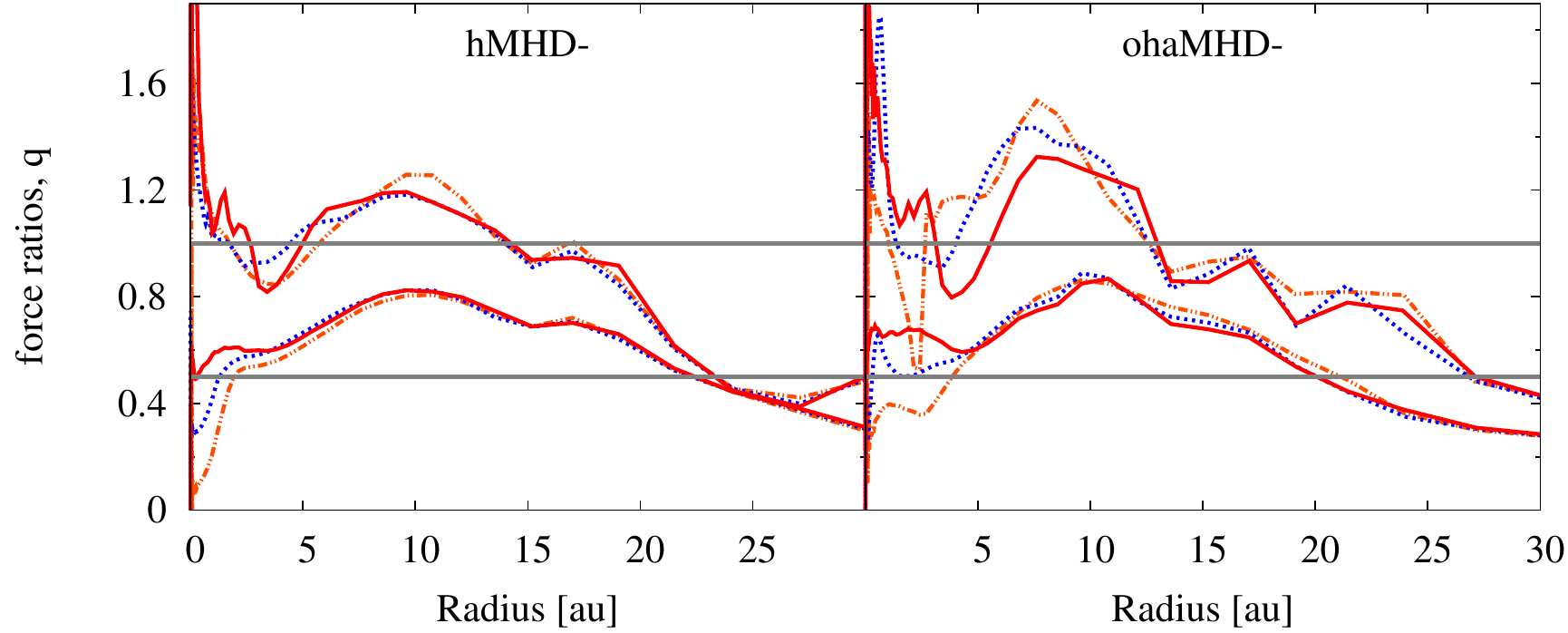}
\caption{The ratio of the sum of the centrifugal and pressure forces to the radial gravitational force is the top set of lines in each panel ($q_1$, Eqn.~\ref{eq:q1}); the ratio of the centrifugal force to the radial gravitational force is the bottom set of lines ($q_2$, Eqn.~\ref{eq:q2}).  The forces are calculated using azimuthal averages of the gas within $20^\circ$ of the mid-plane at the same epochs as in \figref{fig:disc}.  The horizontal lines are reference lines.  Large rotationally supported discs form early in the models that include the Hall effect when the magnetic field and rotation vectors are anti-aligned (bottom panel).  Small discs form near the formation of the stellar core in the models that include the diffusive terms.  Discs do not form for iMHD or hMHD+.} 
\label{fig:disc:q}
\end{figure}

To verify the presence of the disc in hMHD- and ohaMHD-, and to determine if discs have formed in the remaining models, we consider the ratio of centrifugal and pressure forces to the radial gravitational force, 
\begin{equation}
\label{eq:q1}
q_1 = \left| \frac{\frac{v_\phi^2}{r} + \frac{1}{\rho}\frac{\text{d}P}{\text{d}r}}{\frac{GM(r)}{r^2}}\right|,
\end{equation}
and the ratio of centrifugal force to the radial gravitational force,
\begin{equation}
\label{eq:q2}
q_2 = \left| \frac{\frac{v_\phi^2}{r} }{\frac{GM(r)}{r^2}}\right|,
\end{equation}
where $P$ is gas pressure, $M(r)$ is the mass enclosed at radius $r$, and $G$ is Newton's gravitational force constant.  If $q_2 > 0.5$, then the gas is primarily supported by centrifugal force, and we can concluded that a disc exists \citepeg{Tsukamoto+2015oa,Tsukamoto+2015hall,\wbp2018hd}.  The ratios $q_1$ and $q_2$ are shown \figref{fig:disc:q}. 

At the three epochs analysed, $q_2 < 0.5$ for iMHD, confirming that discs do not form in the presence of strong, ideal magnetic fields.  As shown in \figref{fig:disc}, rotationally supported discs clearly form for both hMHD- and ohaMHD-, and they do not undergo significant evolution between their formation during the first core phase and \dtsc{4}.  Gas is rotationally supported out to similar radii (when averaging azimuthally; \figref{fig:disc:q}) for both models, although the semi-major axis is slightly larger for ohaMHD- than hMHD- (\figref{fig:disc}).  It is reasonable that ohaMHD- has a slightly larger semi-major axis than hMHD- since ambipolar diffusion acts to weaken the magnetic field, and larger discs tend to form in weaker fields \citepeg{BateTriccoPrice2014,\wpb2016}; our conclusion that ohaMHD- has a slightly larger disc than hMHD- is in agreement with \citet{Zhao+2021}.

Model hMHD+ never forms a rotationally supported disc.  The Hall effect has caused the gas in the mid-plane to slowly counter-rotate up to a radius of \sm8~au (see \figref{fig:disc:v} and \secref{sec:crotation}).  However, the Hall effect has not spun up the gas enough (in the counter-rotating sense) for a rotationally supported disc for form by the end of the simulation at \dtsc{4}.

The remaining models begin to form discs simultaneously with the formation of the stellar core (\rhoxeq{-4}; \dtsczero).  These discs grow to $r \approx 1$--$4$~au by \dtsc{4}.  Of these models, the disc in aMHD is the largest while ohaMHD+ is the smallest.  The former is a result of ambipolar diffusion weakening the magnetic field, yielding a disc consistent in size with those found in \citet{TomidaOkuzumiMachida2015} and \citet{Tsukamoto+2015oa}.  The latter is a result of the Hall effect extracting some of the angular momentum gained by magnetic diffusion.  The order of the decreasing disc sizes in these models is the same as the order of decreasing angular momentum in the first core: aMHD, oMHD, then ohaMHD+.  

Therefore, all non-ideal MHD effects promote disc formation, except for the Hall effect when the magnetic field and rotation vectors are aligned.  Specifically, the Hall effect will promote the formation of a large disc ($r \gtrsim 20$~au) during the first hydrostatic core phase, while Ohmic resistivity and ambipolar diffusion will promote the formation of a small disc ($r \lesssim 4$~au) as the stellar core forms.

\subsubsection{Bi-modality of discs}
Comparing ohaMHD- to ohaMHD+ and hMHD- to hMHD+ suggests that there may be a bi-modality of discs when the Hall effect is included:  Large, rotationally supported discs form when the magnetic field and rotation vectors are anti-aligned, and small or no discs form when the vectors are aligned.  Thus, our results agree with this previously suggested conclusion \citepeg{Tsukamoto+2015hall,Tsukamoto+2017,WursterPriceBate2016}.  However, it was recently argued by \citet{Zhao+2020,Zhao+2021}\footnote{Note that \citet{Zhao+2020} included Ohmic resistivity and the Hall effect while \citet{Zhao+2021} included ambipolar diffusion as well.} that this bi-modality may not exist since their 2D-axisymmetric simulations formed similar sizes of discs for both the aligned and anti-aligned orientations.

In \citet{Zhao+2020}, when the collapse is dominated by the Hall effect (i.e. an MRN grain distribution with $a_\text{g,min} = 0.03$~$\mu$m), 30-50~au disc formed in the anti-aligned orientation but only the inner 10-20~au remained long-lived.  With the aligned orientation, a counter-rotating disc of 20-40~au formed, but only the inner \sm10~au remained long-lived.  Therefore, they concluded that discs of 10-20~au would persist, independent of the initial orientation.  In \citet{Zhao+2021}, when the collapse is dominated by ambipolar diffusion (i.e. an MRN grain distribution with $a_\text{g,min} = 0.1$~$\mu$m), \sm20~au discs form with prograde rotation, independent of the magnetic field orientation or even the presence of the Hall effect!  Indeed the Hall effect had negligible effect on these simulations.  Therefore, assuming the non-ideal effects are strong enough (i.e. very small grains are removed from the MRN distribution), a rotationally supported disc always forms in \citet{Zhao+2020,Zhao+2021}.

The simulations of \citet{Zhao+2020,Zhao+2021} have the benefit of being evolved considerably longer than those presented here, thus a counter-rotating rotationally supported disc may indeed form later during the evolution of hMHD+.  Therefore, the bi-modality presented here and the literature may be a short-lived feature that is destroyed as the envelope continues to collapse.  Note, however, that \citet{WursterBate2019} ran their simulations longer than presented here, and the bi-modality persisted in their slowly-rotating, strongly magnetised models; all their discs had a prograde rotation.

In summary, a retrograde disc may require very specific conditions to form, and the majority of the discs will rotate in the same sense as the gas from whence it is born.  The bi-modality likely holds at least at early times, but additional 3D global studies are required to determine the long-term persistence of the bi-modality and the sensitivity on the grain size distribution.

%----
\subsection{Outflows}
\label{sec:outflow}
During the gravitational collapse of the cloud, the magnetic field lines get dragged in, pinched and rotated, where the amount of dragging, pinching and rotating depends on the environment and included processes.  This evolution of the magnetic field permits magnetically-launched outflows to form, either as magnetocentrifugal jets \citepeg{BlandfordPayne1982,OuyedPudritz1997} or Poynting-flux-dominated magnetic tower jets \citepeg{ShibataUchida1986,Lyndenbell1996,Ustyugova+2000,Lovelace+2002,NakamuraMeier2004}, where the difference is the distance out to which the magnetic field is dominant \citepeg{Huarteespinosa+2012}.

We have previously shown that strong first core outflows are launched from ideal MHD simulations and aligned non-ideal MHD simulations, and magnetically-launched second core outflows are only launched in ideal MHD simulations or non-ideal MHD simulations with high ionisation rates \citepeg{\wbp2018sd,\wbp2018hd}.  Thermally-launched second-core outflows are launched in non-ideal MHD models only after the formation of the stellar core \textit{if} the cloud core is initially turbulent \citepeg{WursterLewis2020sc}.  Thus, which non-ideal processes are responsible for suppressing outflows?  

\begin{figure} 
\centering
\includegraphics[width=\columnwidth]{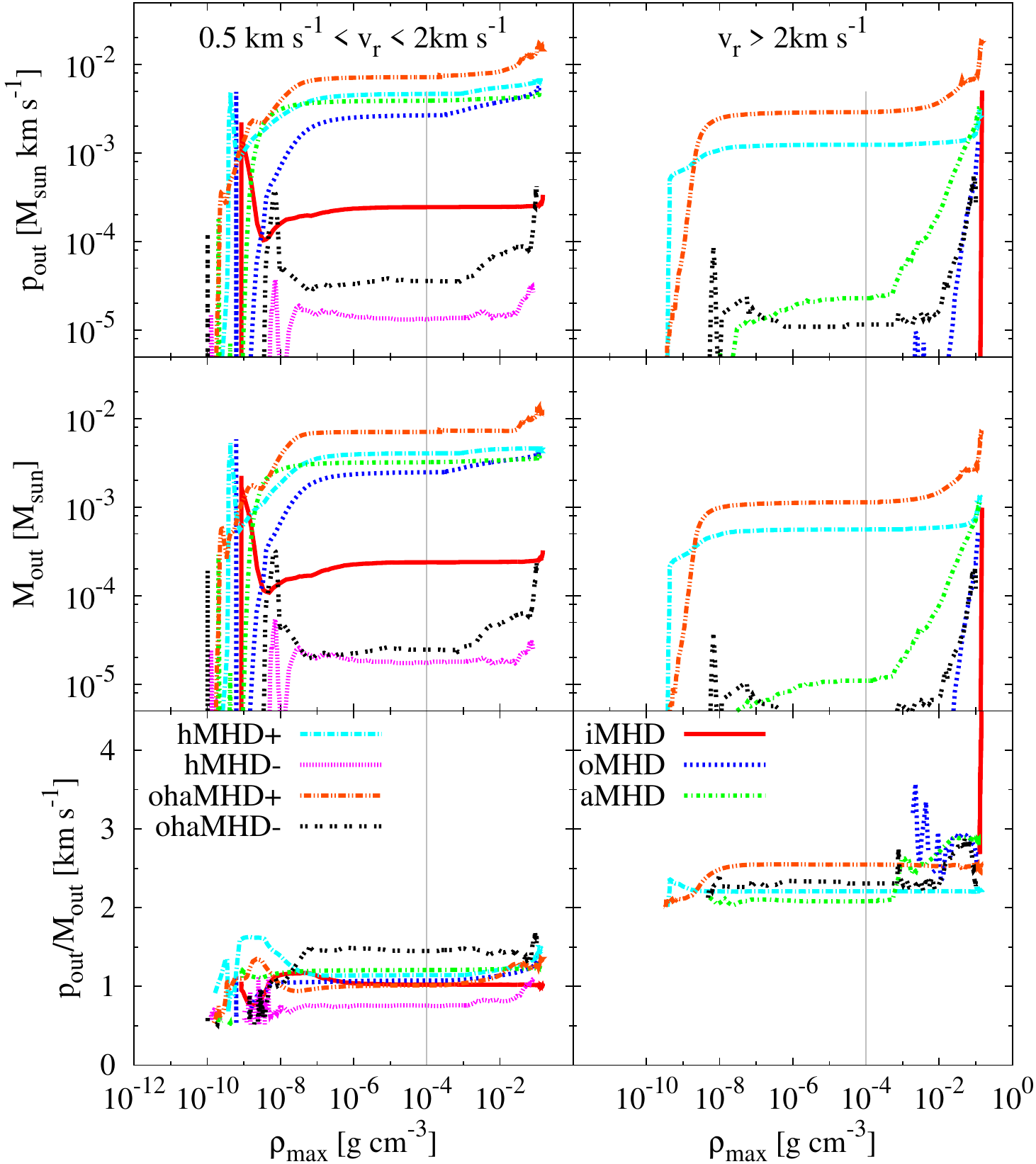}
\caption{The total momentum (top row), total mass (middle row) and average velocity (bottom row) in the slow ($0.5$ \kms$ < v_\text{r} \ < 2$ \kms; left-hand column) and fast ($v_\text{r} > 2$~\kms; right-hand column) outflows. The vertical grey line represents the defined formation density of the stellar core.  Gas is in the outflow if it is 30$^\circ$ above/below the mid-plane and has \rhole{-8}, has $|v_\text{r}|/|v| > 0.5$,  and its radial velocity vector is at least 30$^\circ$ above/below the mid-plane.  The outflowing gas in hMHD- and ohaMHD- is associated with the disc and not a true outflow.  Reasonable first core outflows exist in all models except hMHD- and ohaMHD-; reasonable second core outflows exists in all models that exclude the Hall effect.  Substantial fast first core outflows require the Hall effect and aligned magnetic field and rotation vectors.} 
\label{fig:outflow}
\end{figure}

\figref{fig:outflow} shows the momentum, mass and average outflow velocity of the fast ($v_\text{r} > 2$~\kms) and slow ($0.5 < v_\text{r}/$\kms$ < 2$) outflows.  The value of 2~\kms{} is somewhat arbitrary, but we have tested to ensure that the particular value does not affect our conclusions.  We consider gas that is 30$^\circ$ above/below the mid-plane and has \rhole{-8}, $|v_\text{r}|/|v| > 0.5$,  and whose radial velocity vector is at least 30$^\circ$ above/below the mid-plane to be part of the outflow.  We caution that this criteria permits some material that is in the disc to be categorised as outflow material\footnote{Stricter outflow criteria remove too much outflowing gas and present an inaccurate description of the outflows.}.

This figure does not make the distinction between first and stellar core outflows, although the presence of the stellar core outflow is visible in several of the models.  By comparing the outflows at similar maximum densities, not all models have had a similar length of time to launch the outflows, however, there is no consistent correlation between outflow properties and length of time in the first core phase.  All models contain some outflowing material, however, this outflowing gas in hMHD- and ohaMHD- is primarily associated with the disc.  Models hMHD+ and ohaMHD+ launch the fastest and most massive outflows, indicating that the Hall effect and the aligned orientation of the magnetic field and rotation vectors is required to launch substantial, fast outflows.

\begin{figure*} 
\centering
\includegraphics[width=0.9\textwidth]{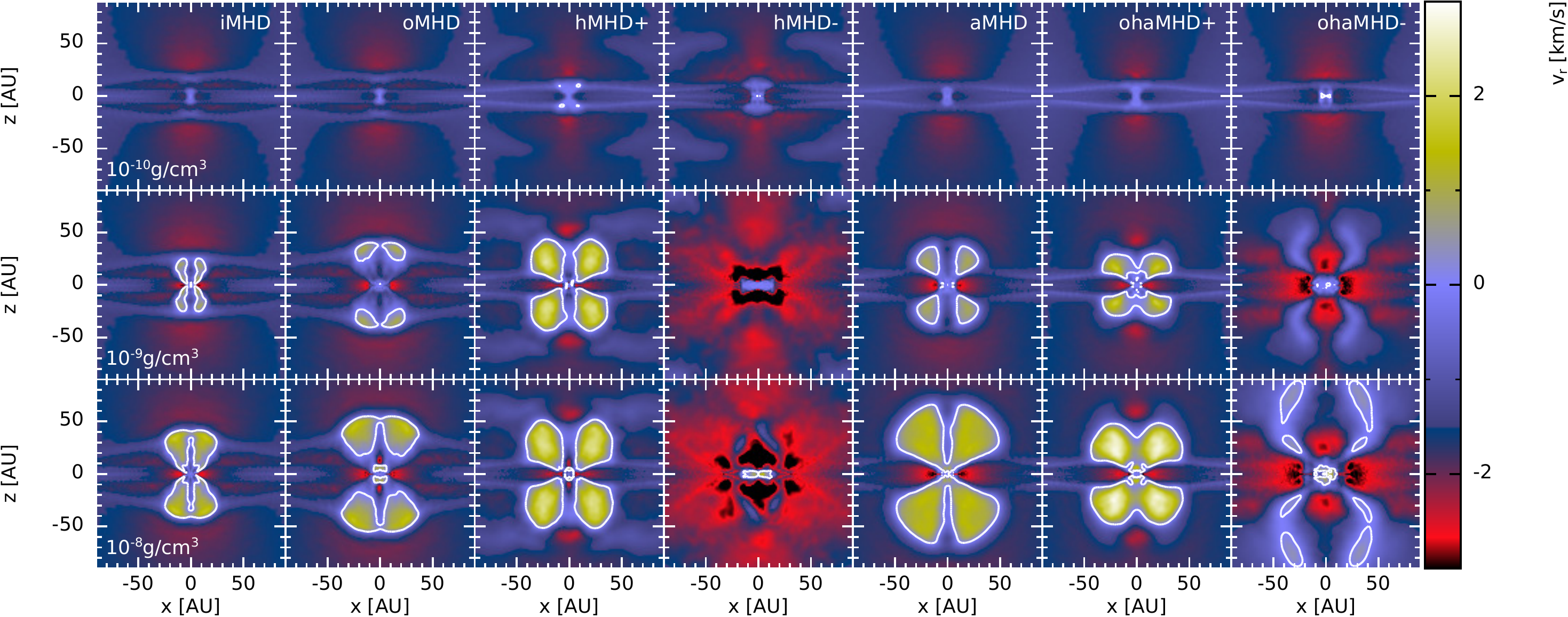}
\includegraphics[width=0.9\textwidth]{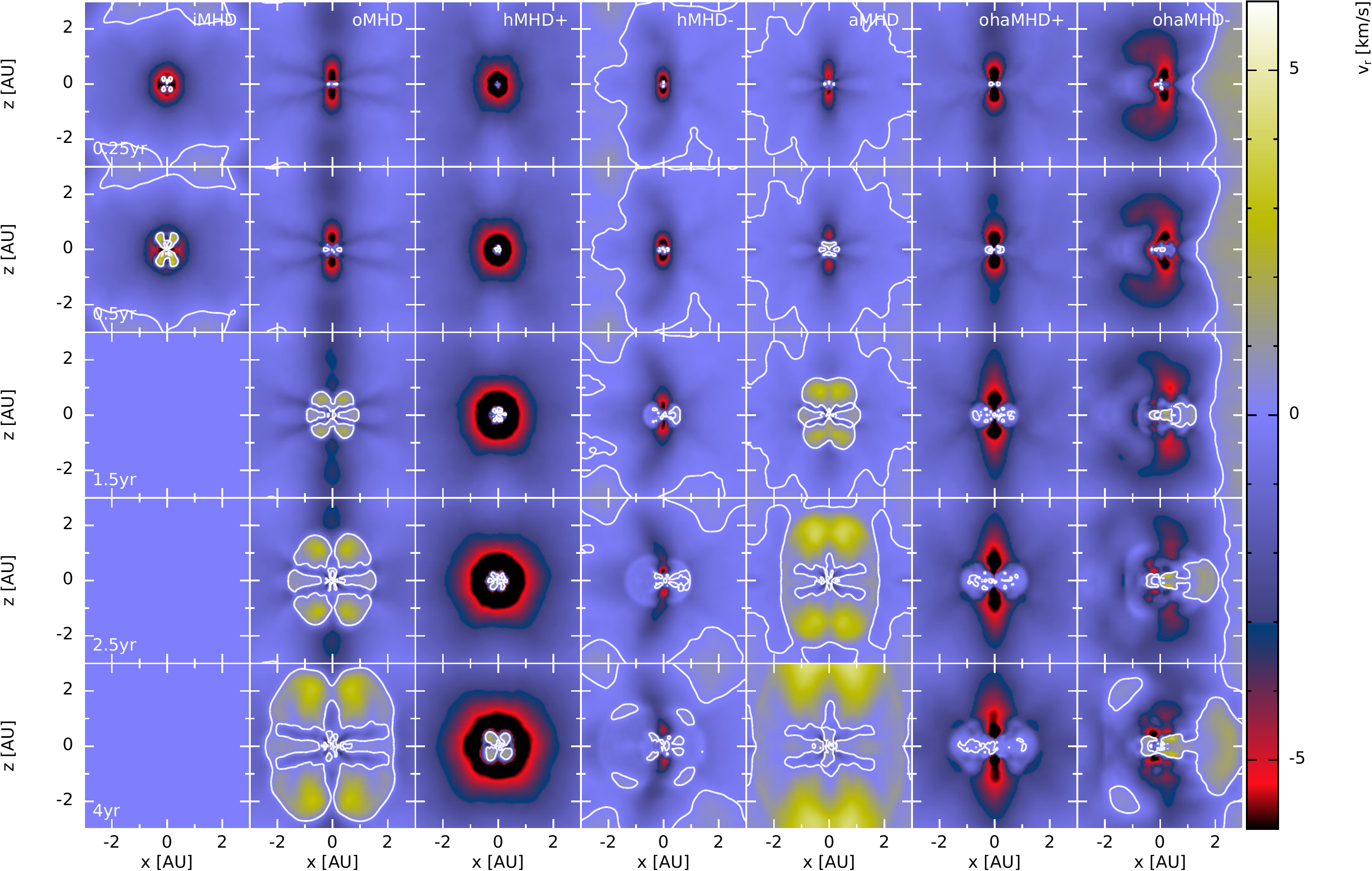}
\caption{Radial velocity slices through the core parallel to the rotation axis throughout the first hydrostatic core phase (top panel) and the stellar core phase (bottom panel).  The contours are at $v_\text{r} = 0$.  Model iMHD ended prior to \dtsc{1.5}, hence the blank frames. The Hall effect plays the dominant role in characterising the outflow.  Fast, massive first core outflows form for the Hall effect with the aligned magnetic field (hMHD+, ohaMHD+), while first core outflows are greatly or totally suppressed for the Hall effect with the anti-aligned magnetic field (hMHD-, ohaMHD-).  The Hall effect with the aligned magnetic field greatly hinders the stellar core outflow, and completely suppresses it in conjunction with the diffusive terms.  The Hall effect with the anti-aligned magnetic field completely suppresses the stellar core outflow.}
\label{fig:outflow:v}
\end{figure*}

\figref{fig:outflow:v} shows the radial velocity at several times during the first and stellar core phases.

\begin{figure*} 
\centering
\includegraphics[width=0.9\textwidth]{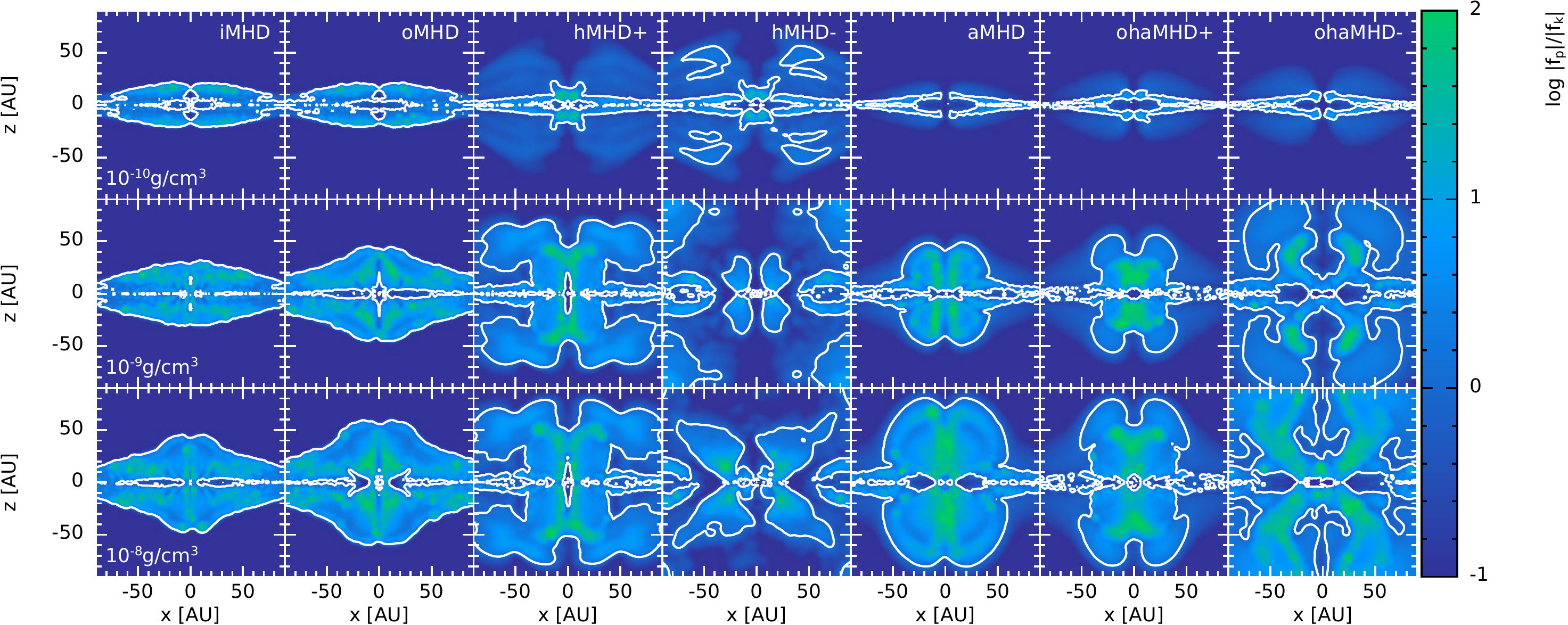}
\includegraphics[width=0.9\textwidth]{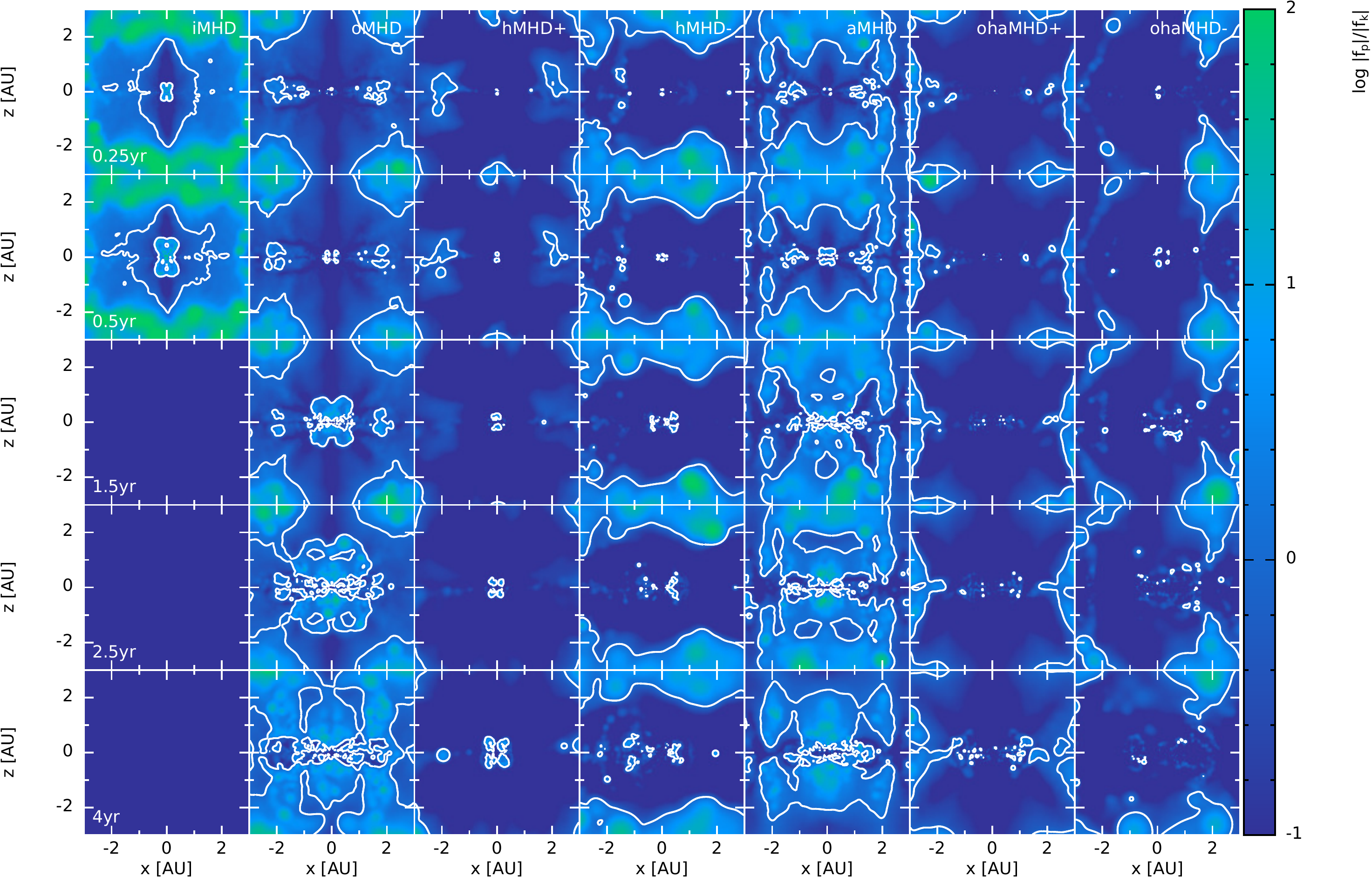}
\caption{The ratio of the vertical component of the Poynting flux $f_\text{P}$ to the vertical component of the kinetic flux $f_\text{k}$ through the core parallel to the rotation axis throughout the first hydrostatic core phase (top panel) and the stellar core phase (bottom panel).  The contours are at $|f_\text{P}|/|f_\text{k}| = 1$.  In the outflows, $|f_\text{P}| > |f_\text{k}|$, indicating that they are still magnetically dominated and have not yet reached the hydrodynamic regime.}
\label{fig:outflow:fpfk}
\end{figure*} 

\subsubsection{First core outflows}
As the first hydrostatic core forms, there is a brief outflow of material rebounding off the core.  This material remains at $r \lesssim 7$~au and quickly dissipates, hence the brief spike in several models as shown in \figref{fig:outflow}.  Immediately after this initial rebound, the first core outflow is launched.   The first core outflow has a fast component for hMHD+ and ohaMHD+, and reasonable ($M \gtrsim 10^{-4}$~\Msun) slow components for all models except hMHD- and ohaMHD-.
In all models with reasonable outflows, the majority of the outflowing gas has velocities of $1 < v_\text{r}/($\kms$) < 2$, which is generally similar to the azimuthal velocity in the mid-plane outside of the disc (see \figref{fig:disc:v}).  

There is no correlation between outflow properties and angular momentum in the first core.  The possible exceptions are anti-correlation of hMHD- and ohaMHD- which have the highest first core angular momentum (and fastest rotating mid-plane) yet the weakest outflows.  Similarly, the only correlation between outflow properties and disc radii is that the models with the weakest outflows have the largest discs; however, the model with the next weakest outflow (iMHD) forms no disc while the remaining models have reasonable outflows and form small discs (or no disc for hMHD+).  There is also no direct relationship between the outflow speed and azimuthal velocity of the mid-plane.  Outflow velocities are greater than azimuthal velocities for hMHD+ and ohaMHD+, similar for iMHD, oMHD aMHD and clearly less than the azimuthal velocity for hMHD- and ohaMHD- which do not form outflows.  

\begin{figure*} 
\centering
\includegraphics[width=0.9\textwidth]{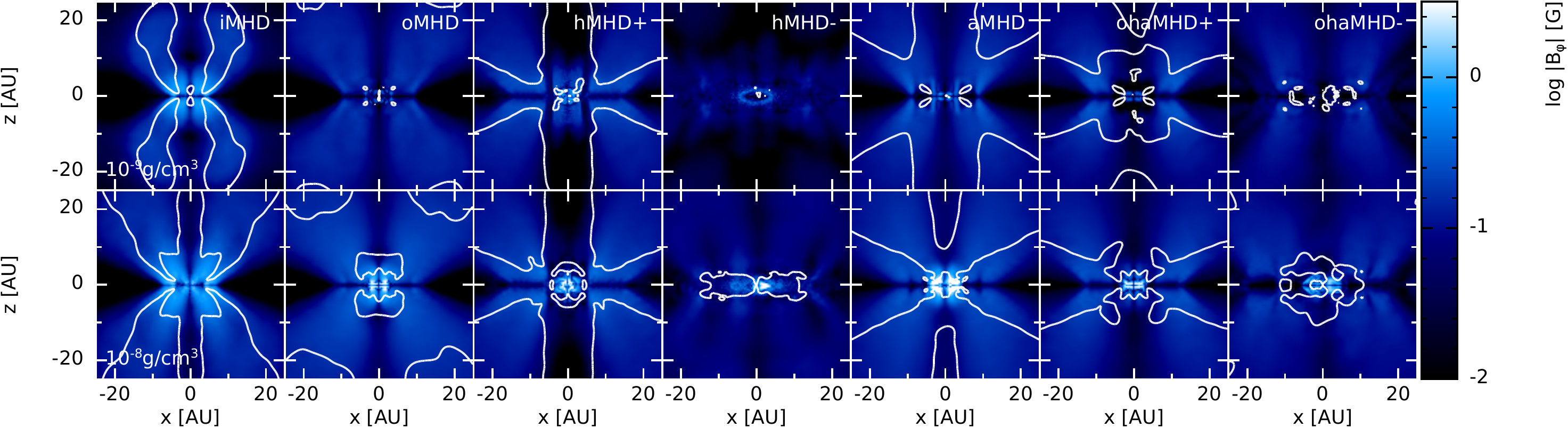}
\includegraphics[width=0.9\textwidth]{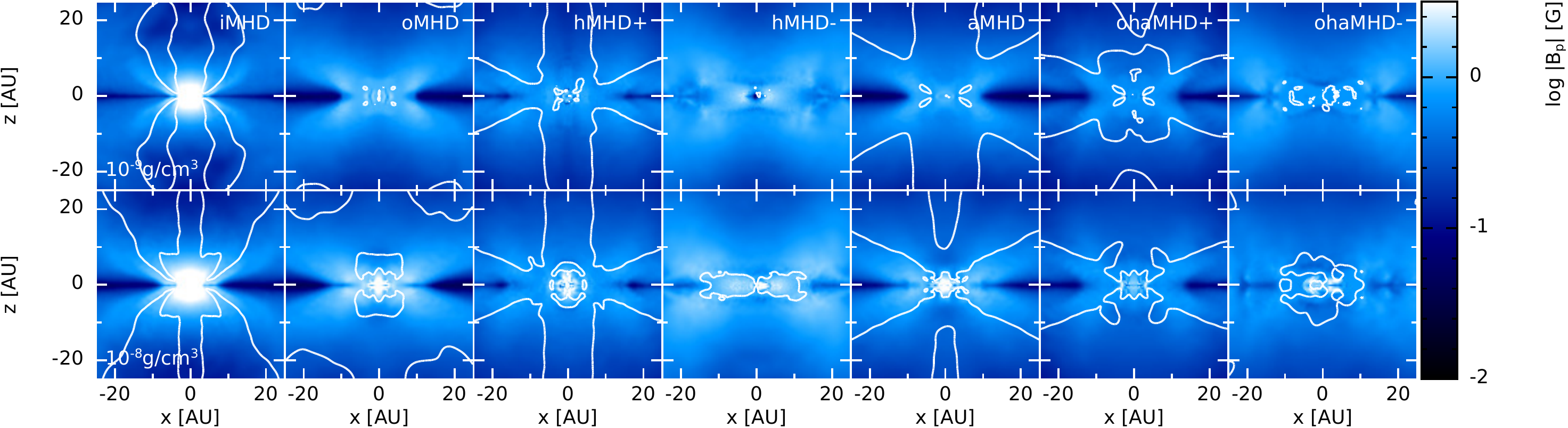}
\includegraphics[width=0.9\textwidth]{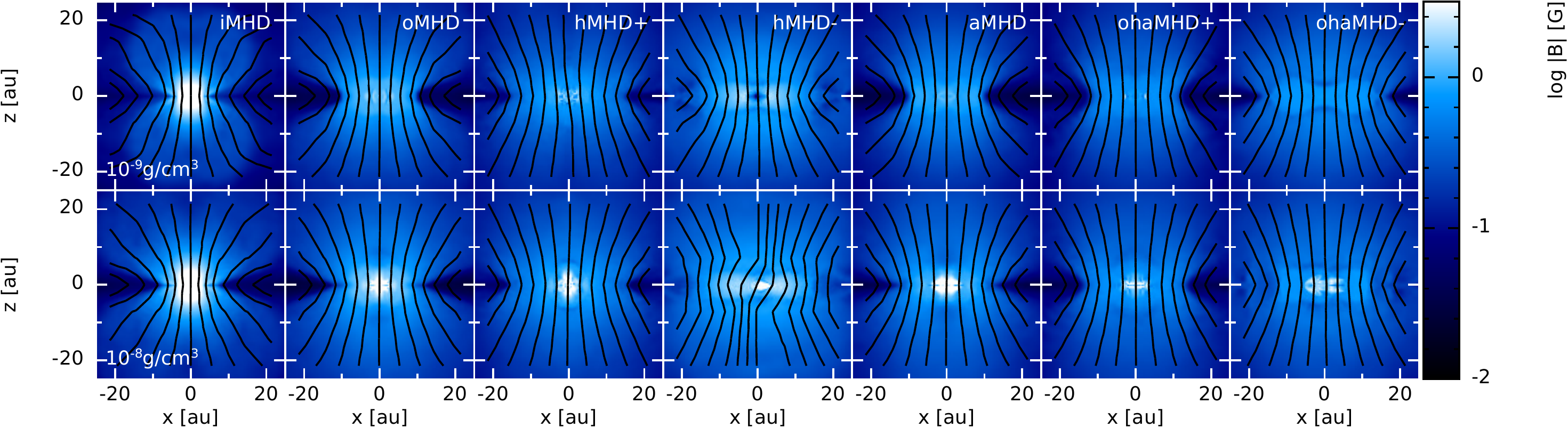}
\includegraphics[width=0.7875\textwidth,trim=0 355 0 0,clip]{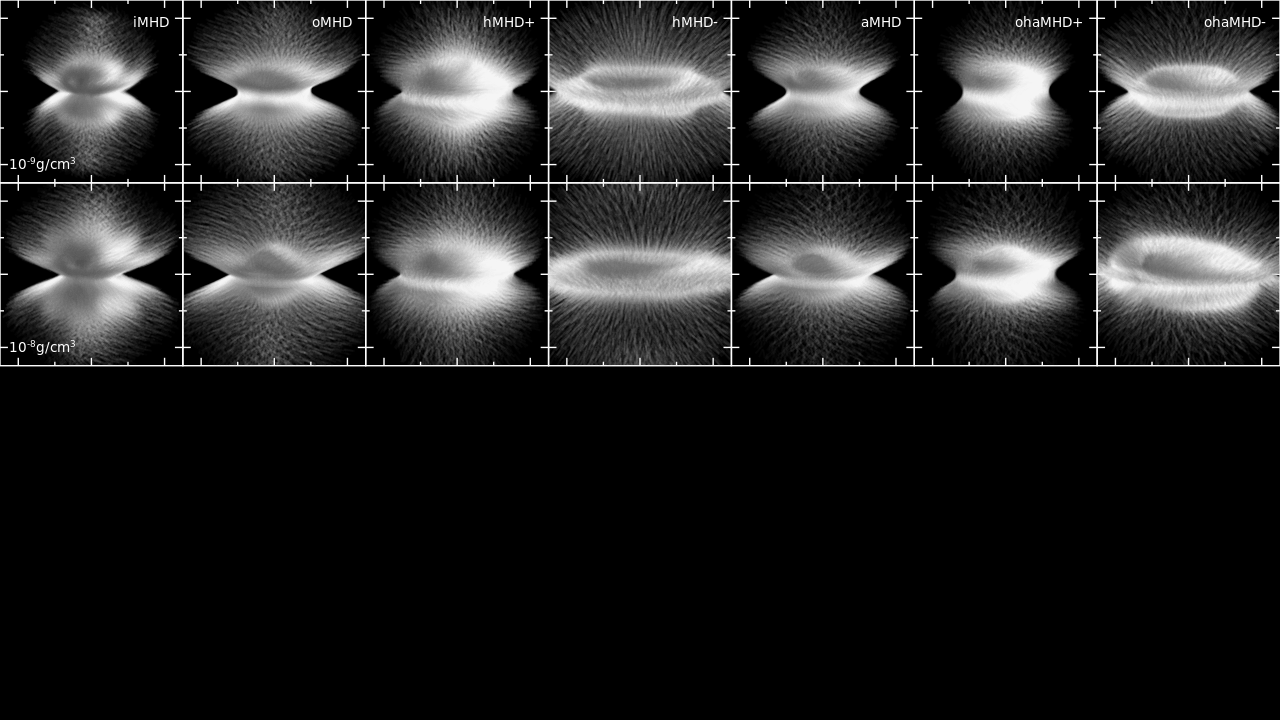}
\caption{The toroidal (top), poloidal (second from top) and total (third) magnetic field strengths in a slice through the first core perpendicular to the rotation axis near the launching region of the first core outflow.  In the top two panels, the contour lines are at $v_\text{r} = 0$, and in the third panel, the streamlines represent the magnetic field.  Since two-dimensional streamlines do not fully capture the geometry of the magnetic field, the bottom panel shows a visualisation of the magnetic field geometry for $0.16 < |\bm{B}|/G < 160$; the box size is the same as the upper panels, but the cores are inclined by $10^\circ$ out of the page.  The field is strongly pinched for iMHD while there is more twisting for the remaining models.  Outflows are launched in models with reasonable toroidal components of the magnetic fields (i.e. not hMHD- and ohaMHD-), and the broadness depends on how `pinched' the magnetic field is.}
\label{fig:outflow:B}
\end{figure*}

From \figsref{fig:outflow}{fig:outflow:v}, it is clear that the Hall effect with the anti-aligned magnetic field is the process that suppresses the first core outflow.   No outflow forms in hMHD- or ohaMHD-.  The Hall effect in this configuration prevents an accumulation of the toroidal component of the magnetic field, as shown in the top panel of \figref{fig:outflow:B}, and without a reasonable toroidal component to the magnetic field, outflows are not launched.  The `outflowing' gas in hMHD- is gas that is puffing up the disc rather than escaping.  In ohaMHD-, the infalling gas is slowing down, most drastically in the X-shaped lobes that contain outflowing gas in the remaining models (except hMHD-) and along the rotation axis.  As this slowdown proceeds in the lobes, the gas spins up (see \figref{fig:counter:vy} and \secref{sec:crotation} below), permitting some gas to reach $v_\text{r} \gtrsim 0$, with the gas at larger radii reaching $v_\text{r} \gtrsim 0$ before the gas closer to the centre of the core.  Thus, the `outflowing' gas in these lobes is a result of a spin-up and not a true outflow.

%iMHD mass during second collapse = 0.000235Msun ; oMHD mass during second collapse = 0.00248Msun
Models iMHD, oMHD, aMHD and hMHD+ all accumulate a reasonable toroidal component of the magnetic field (top panel of \figref{fig:outflow:B}) and launch outflows.  However, the structure of each outflow is different in terms of opening angle, cavity structure along the $z$-axis and the launching regions.  Including Ohmic resistivity does not increase the average velocity of the outflow over iMHD, however there is more mass and hence momentum in the outflow.  Although oMHD remains in the first core phase for \sm30 per cent longer than iMHD, its outflow contains over ten times the mass, indicating that the difference in outflows is simply not due to the length of time in the first core phase.  Unlike iMHD, the gas behind the leading edge of the outflow in oMHD decreases in velocity to slowly fall back onto the core.  Thus, resistive MHD with only Ohmic resistivity will modify the outflow compared to ideal MHD.

Model aMHD launches broad outflows with slightly more mass than that of oMHD.  By the end of the first core phase, the launching region is much closer to the centre of the core than oMHD or even iMHD due to a much stronger toroidal magnetic field component.  This outflow is also broader and contains a wider central cavity.  

When the magnetic field is in the aligned configuration, the Hall effect spins down the gas above/below the core and contributes to the toroidal magnetic field strength, which promotes a fast, massive outflow.  Unlike Ohmic resistivity or ambipolar diffusion, this produces a reasonable mass ($M > 10^{-4}$~\Msun{}) of both fast and slow moving gas within the first core outflow.  Therefore, to generate a fast first core outflow, the Hall effect is required with the magnetic field and rotation vectors aligned.

The difference in structure is a result of how the non-ideal processes reshape the magnetic field.   Although the field typically remains overall strongly poloidal (second panel of \figref{fig:outflow:B}), the magnetic field lines are much more `pinched' for iMHD than the models that include a non-ideal process (third and fourth panels of \figref{fig:outflow:B}).  The reduced pinching, as in hMHD+ or aMHD is not so much that it prevents outflows from forming \citepeg{BlandfordPayne1982}, but it does broaden the outflow insomuch as the escaping gas follows the less-pinched field lines.  This results in a more distinct central cavity in hMHD+ and aMHD than in iMHD.

The cumulative effect of the diffusive terms and Hall effect in the aligned orientation (i.e., ohaMHD+) yields the fastest and most massive outflow with the most momentum in our suite.  This is a combined result of the diffusive terms reducing the pinching of the magnetic field to allow for a broader outflow and the Hall effect contributing to the toroidal component of the magnetic field.  This is a clear departure from both ideal and resistive (Ohmic only) MHD, reinforcing the necessity of all non-ideal terms and a clear understanding of the magnetic field geometry.

The cumulative effect of the diffusive terms and Hall effect in the anti-aligned orientation (i.e., ohaMHD-) yields no outflow, as discussed above.  Although the diffusive terms slow down the infalling gas in the X-shaped lobes, they are clearly not strong enough to overcome the Hall effect to permit a first core outflow to be launched.

Following the analysis in \citet{Huarteespinosa+2012} and \citet{BateTriccoPrice2014}, we compare the vertical component of the Poynting flux,
\begin{equation}
f_\text{P} = \left[ \bm{B} \times \left( \bm{v} \times \bm{B} \right) \right]_\text{z},
\end{equation}
where $B$ and $v$ are the magnetic field and velocity vectors, respectively, to the vertical component of the kinetic flux,
\begin{equation}
f_\text{k} = \frac{1}{2}\rho v^2v_\text{z},
\end{equation}
where $\rho$ is gas density and $v_\text{z}$ is the vertical component of the velocity.
For the outflows that form in our suite,  $|f_\text{P}| > |f_\text{k}|$, indicating that the bulk of the outflows are Poynting flux dominated, meaning that they are still magnetically dominated and have not yet reached the hydrodynamic regime; see top panel of \figref{fig:outflow:fpfk}.  Therefore, independent of which non-ideal processes are included, if first core outflows are launched, they are magnetic tower jets.

Therefore, the structure of the first core outflow results primarily from the inclusion of the Hall effect, and is dependent on the magnetic field geometry: the Hall effect in the aligned orientation yields fast, massive outflows while the Hall effect in the anti-aligned orientation suppresses outflows.

\subsubsection{Stellar core outflows}
\begin{figure*} 
\centering
\includegraphics[width=0.9\textwidth]{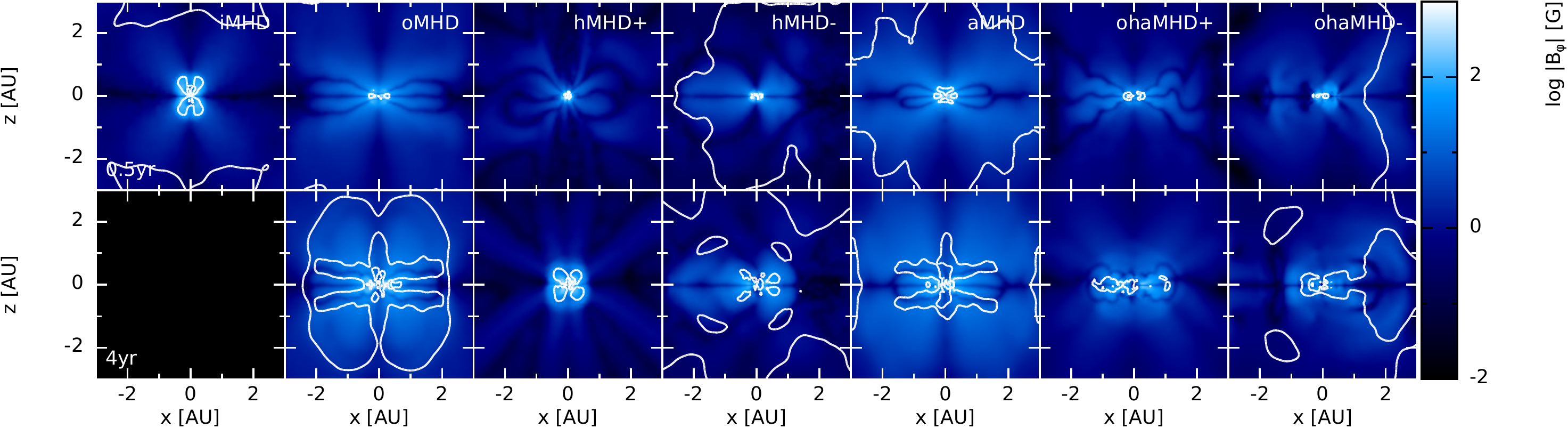}
\includegraphics[width=0.9\textwidth]{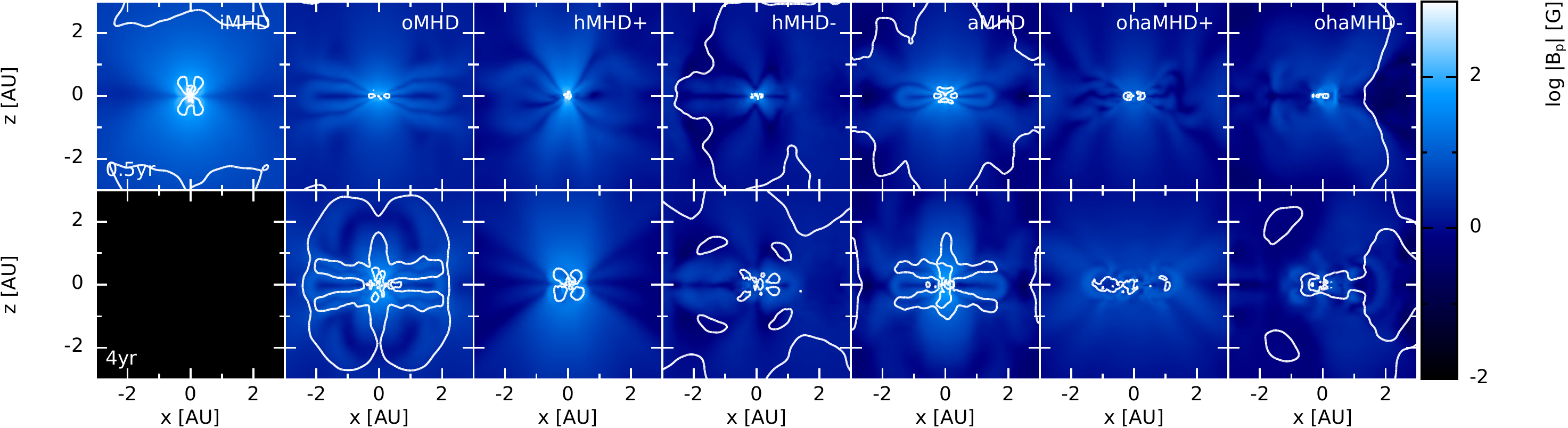}
\includegraphics[width=0.9\textwidth]{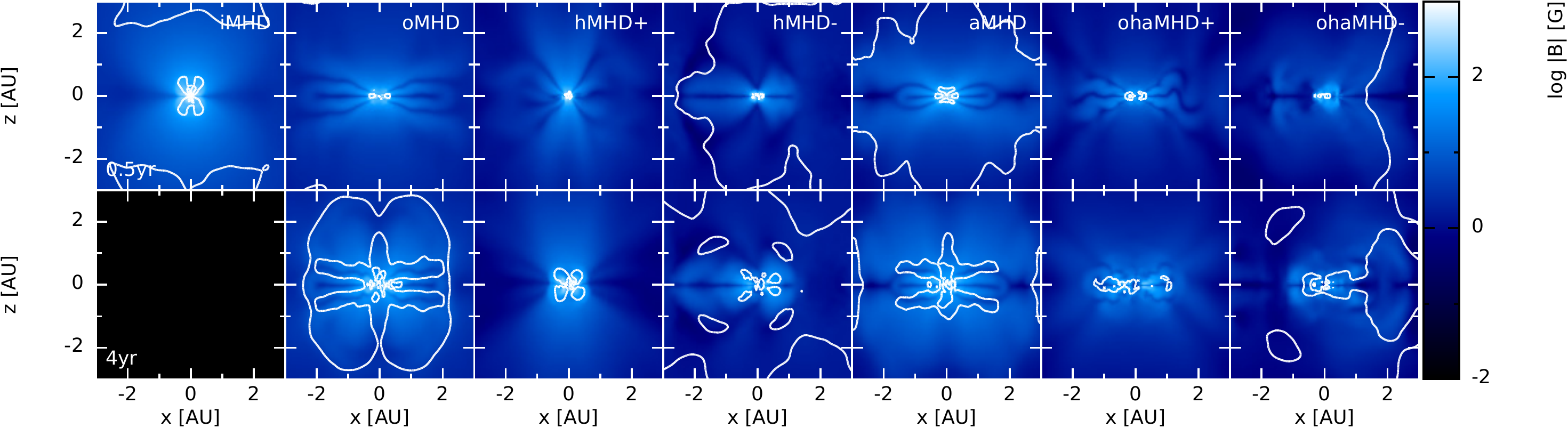}
\includegraphics[width=0.9\textwidth]{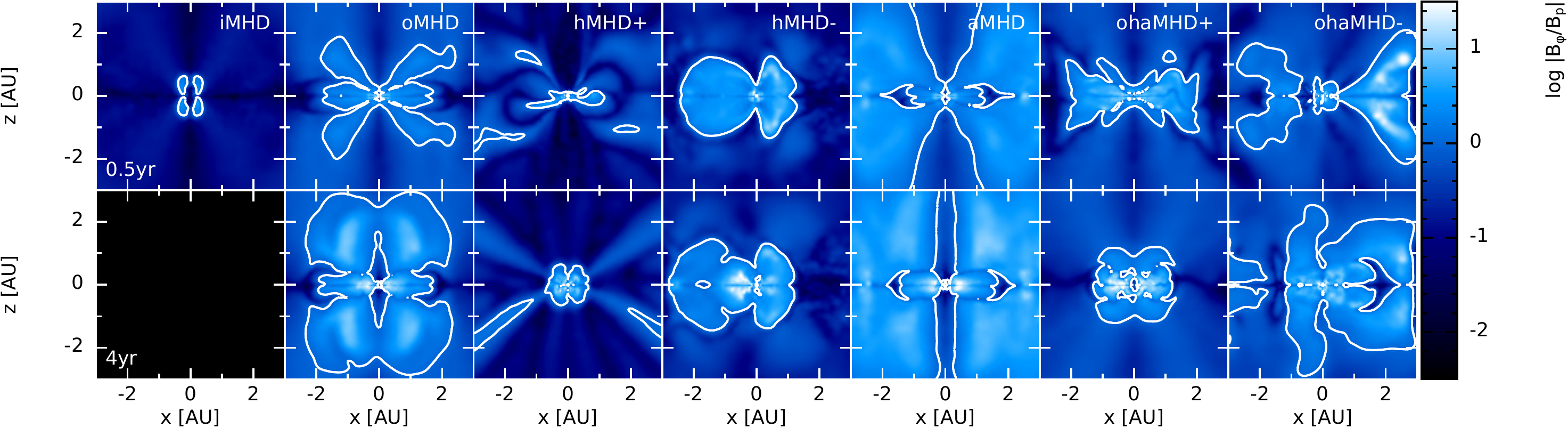}
\includegraphics[width=0.7875\textwidth,trim=0 432 272 0,clip]{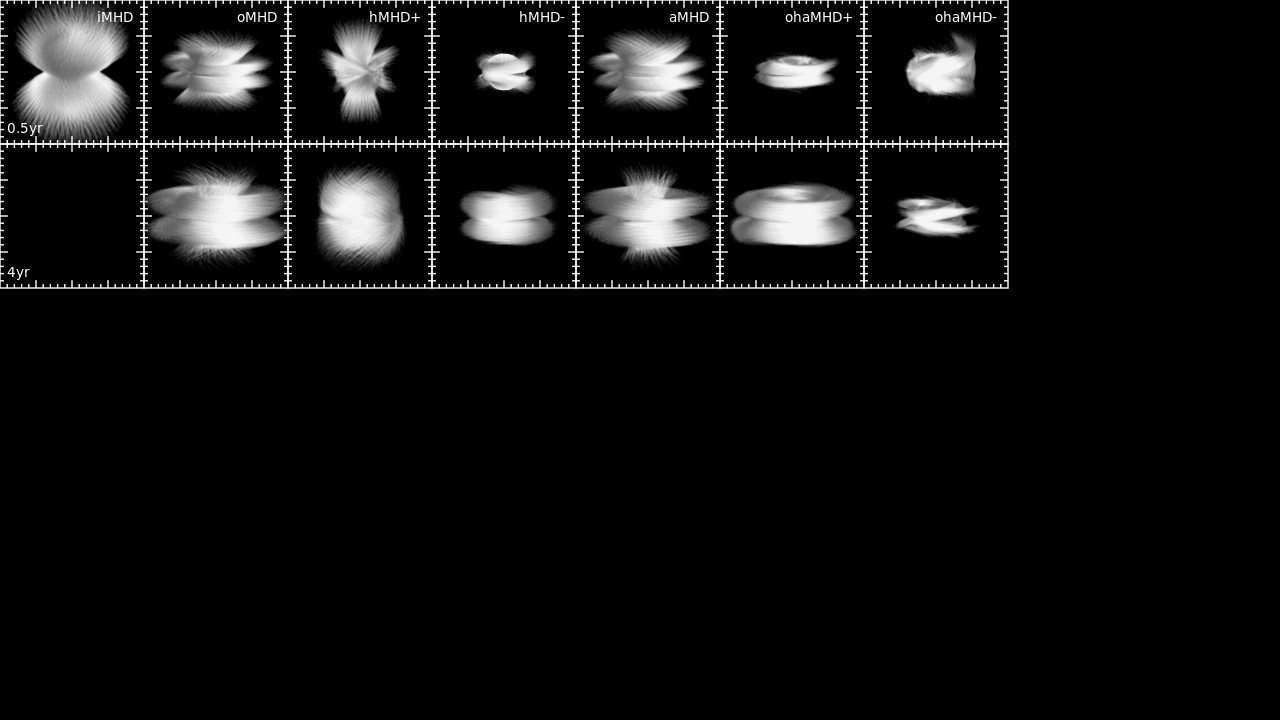}
\caption{The toroidal (top), poloidal (second from top) and total (third) magnetic field strengths in a slice through the stellar core perpendicular to the rotation axis near the launching region of the stellar core outflow.  The fourth panel shows the ratio of the toroidal and poloidal magnetic field components.   In the top three panels, the contour lines are at $v_\text{r} = 0$, and in the fourth panel the contour is at $|B_\phi/B_\text{p}| = 1$.  The bottom panel shows a visualisation of the magnetic field geometry for $36 < |\bm{B}|/G < 3.6\times10^4$; the box size is the same as the upper panels, but the cores are inclined by $10^\circ$ out of the page.  Model iMHD ended prior to \dtsc{4}, hence the blank frames.  The field is strongly pinched for iMHD while the remaining models are dominated by strong toroidal components; the outflows are dominated by the toroidal magnetic field.  The diffusive terms broaden and delay the outflow, while the Hall effect completely suppresses magnetically launched outflows, both on its own an in conjunction with the diffusive terms.  }
\label{fig:outflow:Bsc}
\end{figure*}

Once the stellar core has formed at \rhoxeq{-4}, the density continues to rapidly increase to \rhoxapprox{-1} in just over a month for iMHD but the increase slows for the non-ideal models \citepeg{BateTriccoPrice2014,\wbp2018sd,\wbp2018hd}.  Within a month, iMHD launches a fast stellar core outflow, as seen in \figsref{fig:outflow}{fig:outflow:v}.  Although the model is evolved for only \sm8~mo after stellar core formation, it is the fastest stellar core outflow in our suite and contains the most momentum.  At \dtsc{0.5}, the maximum outflow velocity $v_\text{r} \sim 10$~\kms{}, which is much faster than the maximum azimuthal velocity at that time of $v_\phi \sim 2$~\kms{} (cf. \figref{fig:disc:v}).  Model iMHD also has the strongest magnetic field in the stellar core (see also \figref{fig:BVrho}), and a strong toroidal component in the outflow that is stronger than the poloidal component (fourth panel of \figref{fig:outflow:Bsc}).  

When all three non-ideal MHD terms are included -- with either aligned or anti-aligned magnetic fields -- stellar core outflows are suppressed \citepeg{\wbp2018sd,\wbp2018hd}.
There is an increase in outflowing material (cf. \figref{fig:outflow}), however, this additional material is in the first core outflow (ohaMHD+) or associated with the disc (ohaMHD-).   But which term is primarily responsible for suppressing this outflow? 

Stellar core outflows are launched \sm1~yr after the formation of the stellar core when only the diffusive terms are included (i.e. oMHD and aMHD).  The magnetic field strength in and near the core of these models is a few orders of magnitude lower than in iMHD.  The weaker magnetic field and a stronger toroidal component compared to iMHD permits a broader outflow to be launched, and the weaker field results in its delay.  The stellar core outflows in oMHD and aMHD are launched at similar times with similar broadness (bottom panel of \figref{fig:outflow}), however the outflow in aMHD is slightly faster given the initially stronger toroidal component of the magnetic field in the surrounding gas.  Similar to iMHD, there is a reasonable contribution from the gas pressure, thus both magnetic pressure (i.e. $\left| \bm{J} \times \bm{B}\right|_\text{z}/\rho$) and thermal pressure (i.e. $\left| \text{d}P/\text{d}r\right|/\rho$) contribute to these stellar core outflows.  However, these outflows remain magnetically dominated (i.e. $|f_\text{P}| > |f_\text{k}|$; see bottom panel of \figref{fig:outflow:fpfk}).  Thus, including Ohmic resistivity or ambipolar diffusion will delay the launching of the outflow and change its structure compared to iMHD, but they do not change the underlying physics of the stellar core outflow.

The stellar core outflow in hMHD- is completely suppressed due to the weak magnetic field strength and the toroidal component of the magnetic field only reasonably existing in the disc.

In hMHD+, the magnetic field in the core is only slightly weaker than in oMHD and aMHD (cf. \figref{fig:BVrho}).  A weak outflow is launched, which contains a strong poloidal field, a stronger toroidal field, and similar magnetic and thermal pressures.  By 4~yr after stellar core formation, this outflow extends less than one au, which is several times less extended than the outflows in oMHD and aMHD at a similar time.  Model hMHD+ has a unique radial velocity structure around the core.  The Hall effect in this model has efficiently transported angular momentum outward permitting a rapid infall ($v_\text{r} < -6$~\kms) of gas around the core; since this slow outflow is colliding with the rapid infall, the infall may contribute to the outflow's short extent.  In the remaining models, the gas is typically outflowing into slowly moving gas ($v_\text{r} \sim 0$), thus its progress is less hindered.

When all three terms are included, the central magnetic field strength is weaker than when only one non-ideal term is included (cf. \figref{fig:BVrho}), although this becomes less clear when considering ohaMHD- and its $m=2$ instability.  Moreover, the toroidal component of the magnetic field of ohaMHD$\pm$ is much weaker than in oMHD or aMHD; comparing oMHD and aMHD with hMHD$\pm$ suggests that this weaker toroidal component is a direct result of the Hall effect.  Therefore, the Hall effect (aligned and anti-aligned) has a greater influence on the magnetic field strength and geometry than the diffusive terms, and the stellar core outflows are completely suppressed in ohaMHD$\pm$.  In ohaMHD+, the diffusive terms weaken the magnetic field enough that a small rotationally-supported disc forms (cf. \secref{sec:discs}) and the small outflow as seen in hMHD+ is absent.  

In summary, the diffusive terms delay the launching of the stellar core outflow and broaden its structure compared to ideal MHD, but they do not suppress its formation.  
The Hall effect completely suppresses the launching of outflows (anti-aligned) or permits the launching of only small, slow outflows (aligned).  In conjunction with the diffusive terms, the Hall effect completely suppresses all stellar core outflows, independent of initial magnetic field orientation.

%----
\subsection{Counter-rotation}
\label{sec:crotation}
The vector evolution of the magnetic field resulting from the Hall effect induces a rotational velocity, $v_\text{Hall}$ \citepeg{KrasnopolskyLiShang2011,LiKrasnopolskyShang2011,BraidingWardle2012acc,Tsukamoto+2015hall,Tsukamoto+2017,\wpb2016}; if the cloud is initially rotating, then $v_\text{Hall}$ will contribute to the azimuthal velocity to spin up or spin down the gas (cf. \figref{fig:disc:v}).  When the gas is spun up in the anti-aligned orientation, the magnetic breaking catastrophe is prevented \citepeg{Tsukamoto+2015hall,Tsukamoto+2017,\wpb2016,\wbp2018hd}.  
One frequent consequence of this spin up/down is the formation of counter-rotating regions. 

\begin{figure*} 
\centering
\includegraphics[width=0.9\textwidth]{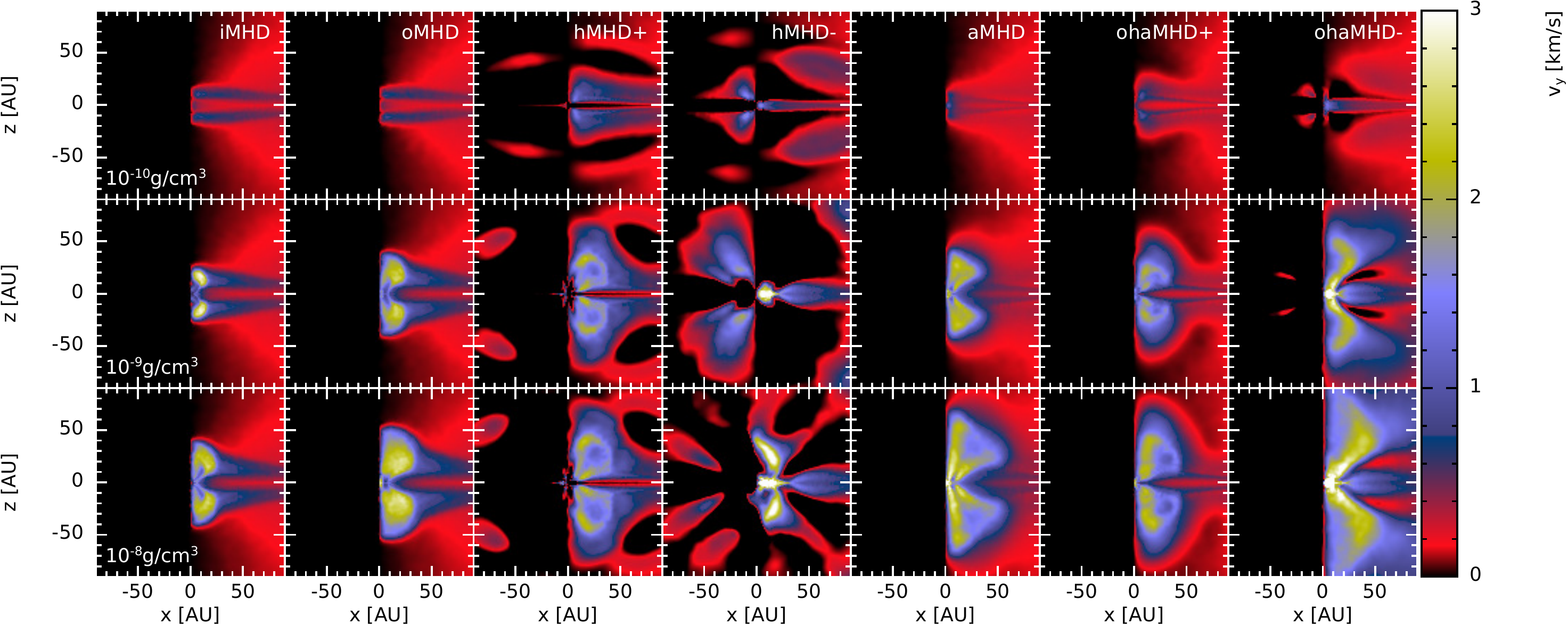}
\includegraphics[width=0.9\textwidth]{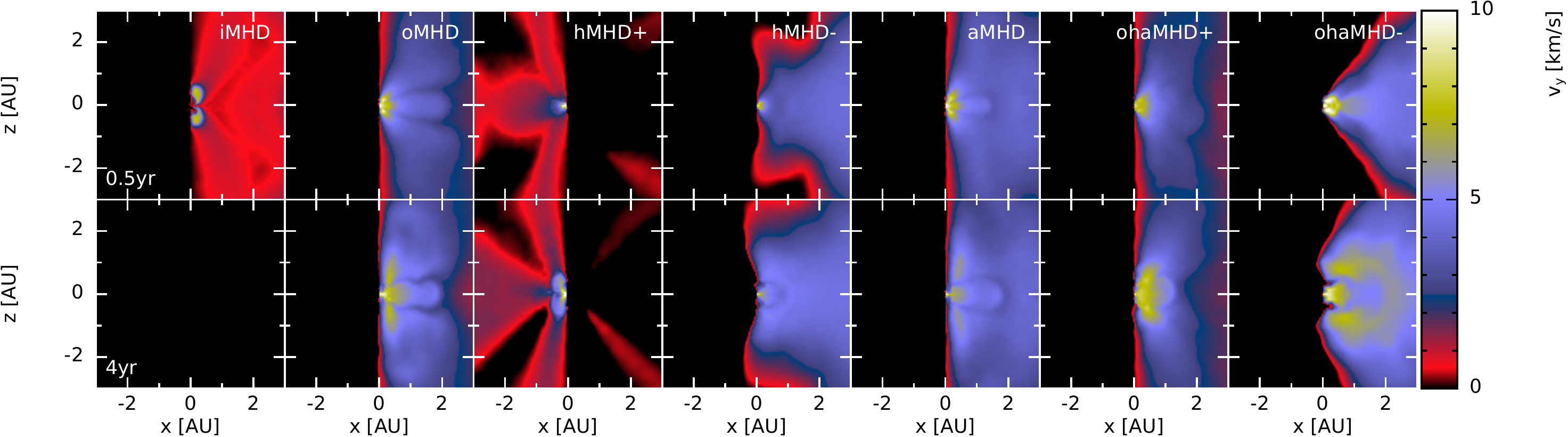}
\caption{$y$-velocity slices through the core parallel to the rotation axis throughout the first hydrostatic core (top panel) and the stellar core (bottom panel) phases.  We intentionally plot only $v_\text{y} > 0$ to highlight the counter-rotating gas.  Model iMHD ended prior to \dtsc{4}, hence the blank frame.  Counter-rotating regions only form in models with the Hall effect.  A counter-rotating pseudo-disc forms in hMHD+, with persistent counter-rotating pockets, while large but transient counter-rotating pockets form in hMHD-.  The diffusive terms weaken the magnetic field to prevent (ohaMHD+) or substantially reduce (ohaMHD-) the counter-rotating regions.}
\label{fig:counter:vy}
\end{figure*}

\figref{fig:counter:vy} shows $v_\text{y} > 0$ during the first and stellar core phases; the cloud's initial rotation is $v_\text{y} > 0$ for $x > 0$.  
As expected, the models that exclude the Hall effect have smooth and predictable rotation profiles.  As the envelope collapses, the gas above/below the mid-plane slowly spins up due to conservation of momentum.  A natural process to extract this increased angular momentum near the centre of the core is outflows (cf. \secref{sec:outflow}).  Indeed, rotating first and second core outflows are launched, which increases rotation near the rotation axis.  In these models, all rotation is in the same direction as the initial rotation -- there is no counter-rotating gas.  

The rotation profile in the models that include the Hall effect (with either magnetic field orientation) is more complex.  In these models, the sign of the Hall effect is negative everywhere, except in the centre of the core \citep{Wurster2021}, thus the Hall effect affects the rotational velocity in the same sense nearly everywhere.   Naturally, the effect is stronger in the denser regions with the stronger magnetic fields.  Since angular momentum is a conserved quantity, any notable modification of the rotational velocity by the Hall effect in one region must be countered by transporting angular momentum to/from another region.  

In the models that include the Hall effect and the anti-aligned magnetic field (i.e., hMHD- and ohaMHD-), $v_\text{Hall}$ contributes constructively to $v_\phi$, spinning up the gas to form a rotationally supported disc (cf. \figrref{fig:disc}{fig:disc:q} in \secref{sec:discs}).  This increased rotation slows down the infall and ultimately prevents the magnetic braking catastrophe.  To spin up the disc, the Hall effect extracts angular momentum from the envelope, spinning down those regions.  Indeed, enough angular momentum is extracted from these regions that they begin to counter-rotate!

In ohaMHD-, the spin-up compared to iMHD results from the diffusive terms weakening the magnetic field and from the Hall effect.  This joint effort results in small counter-rotating pockets that are transient and dissipate before the end of the first core phase.   Larger counter-rotating pockets are seen in the similar model of \citet{Tsukamoto+2015hall}, suggesting that the size and longevity of the pockets may be dependent on the initial rotation and magnetic field strength.\footnote{The model in \citet{Tsukamoto+2015hall} was initialised with a stronger magnetic field strength and faster initial rotation than ohaMHD-.}

Model hMHD- has no diffusive terms, thus the resulting spin-up compared to iMHD is from the Hall effect alone.  This requires the extraction of more angular momentum from the envelope than in ohaMHD-, resulting in much larger counter-rotating pockets that persist up to \rhoxeq{-9}. By this time, the disc has formed and is already rotating at near Keplerian speeds, thus the Hall effect can no longer spin up the mid-plane gas.  The infalling gas above/below the disc near the rotation axis, however,  is spun up, as seen in the third row of \figref{fig:counter:vy}.  As the envelope continues to collapse, the new gas brings with it prograde angular momentum to slowly dissipate the counter-rotating pockets.  Although these pockets are dissipating, they remain at the end of the simulation at \dtsc{4}.  As with ohaMHD-, it is possible that these pockets too will dissipate over longer timescales.  

The gas in ohaMHD+ is spun down by the Hall effect, however, no counter-rotating pockets form due to the diffusive terms weakening the magnetic field.

During the first core phase, the gas in the mid-plane of hMHD+ is spun down such that it becomes counter-rotating, and a counter-rotating pseudo-disc\footnote{The pseudo-disc is disc-shaped, but not rotationally supported.} forms; additional counter-rotating pockets form at $x \approx \pm70$~au.  Unlike in the anti-aligned models, these counter-rotating pockets appear persistent, although their longevity is beyond the scope of this study.  After the formation of the stellar core at \rhoxeq{-4} -- which itself is counter-rotating -- a small, counter-rotating stellar core outflow is launched (see \secref{sec:outflow}).  Thus, in this model, the Hall effect has completely altered the rotational profile of the stellar core and surrounding gas.  

We explicitly note that these counter-rotating pockets (and those in \citealp{Tsukamoto+2015hall,Tsukamoto+2017} and \citealp{WursterPriceBate2016}) formed in simulations where the core was initialised with solid-body rotation.  We also note that they appear to be dependent on initial rotation rate \citep[e.g. comparing to][]{Tsukamoto+2015hall}, magnetic field alignment with the rotation axis \citepeg{Tsukamoto+2017} and magnetic field strength (e.g. comparing hMHD$\pm$ to ohaMHD$\pm$).  Thus, these well-defined (albeit mostly transient) features may be a result of the idealised initial conditions and sensitive to these conditions.  If the magnetic field strength is weaker, or if the gas was initially turbulent, then these pockets may not form at all!  Therefore, to determine the robustness and longevity of counter-rotating pockets, we must examine simulations with less idealised initial conditions, which is out of the scope of this study.  

%----
\subsection{Magnetic walls}
\label{sec:wall}
\citet{TassisMouschovias2005b,TassisMouschovias2007a,TassisMouschovias2007b} characterise a `magnetic wall' as a rapid steepening of the magnetic field strength over a narrow radius.  Similar to \citet{TomidaOkuzumiMachida2015}, our models that include some or all of the non-ideal processes show a steepening of the magnetic field, but it not as pronounced as in (e.g.) \citet{TassisMouschovias2005b}.  However, as we have previously discussed in \citet{\wbp2018ff} in regards to ohaMHD+, magnetic flux piles up in a torus at \sm1-3~au from the centre of the core.  The strongest magnetic field is in this torus rather than coincident with the maximum density, and we refer to this torus as the `magnetic wall.'   While both walls have the same physical manifestation, our walls have a finite thickness where the magnetic field strengths is lower immediately on either side of the wall.

The first hydrostatic core has temperatures of $T \gtrsim 10^3$~K, meaning that thermal ionisation is becoming more important than ionisation from cosmic rays when computing the effect of non-ideal MHD \citep{Wurster2016}.  Although the non-ideal MHD coefficients are decreasing as the first core is becoming more ionised, they still contribute to the evolution of the magnetic field, permitting the wall to form within the hot first hydrostatic core.  If the region (somehow) remained cooler and less ionised, then it is reasonable to expect a stronger wall would form, as in \citet{TassisMouschovias2005b}.

During the formation of this wall in ohaMHD+ during the first core phase, the location of the maximum magnetic field strength decouples from the maximum density (recall \figref{fig:BVrho}).  The magnetic field strength in the wall is a few times higher than the central field strength and remains at these levels during the second collapse phase.  Although the magnetic wall persists, the radius of the maximum magnetic field strength decreases after stellar core formation, but the maximum magnetic field strength never re-couples with the central field strength; indeed, the maximum field strength becomes \sm10-100 times higher than the central field strength.  Using ohaMHD+, the values of the maximum and central field strengths permitted us to conclude that magnetic fields in low-mass stars are generated by a dynamo and are not fossil fields \citep{\wbp2018ff}.

\begin{figure*} 
\centering
\includegraphics[width=0.9\textwidth]{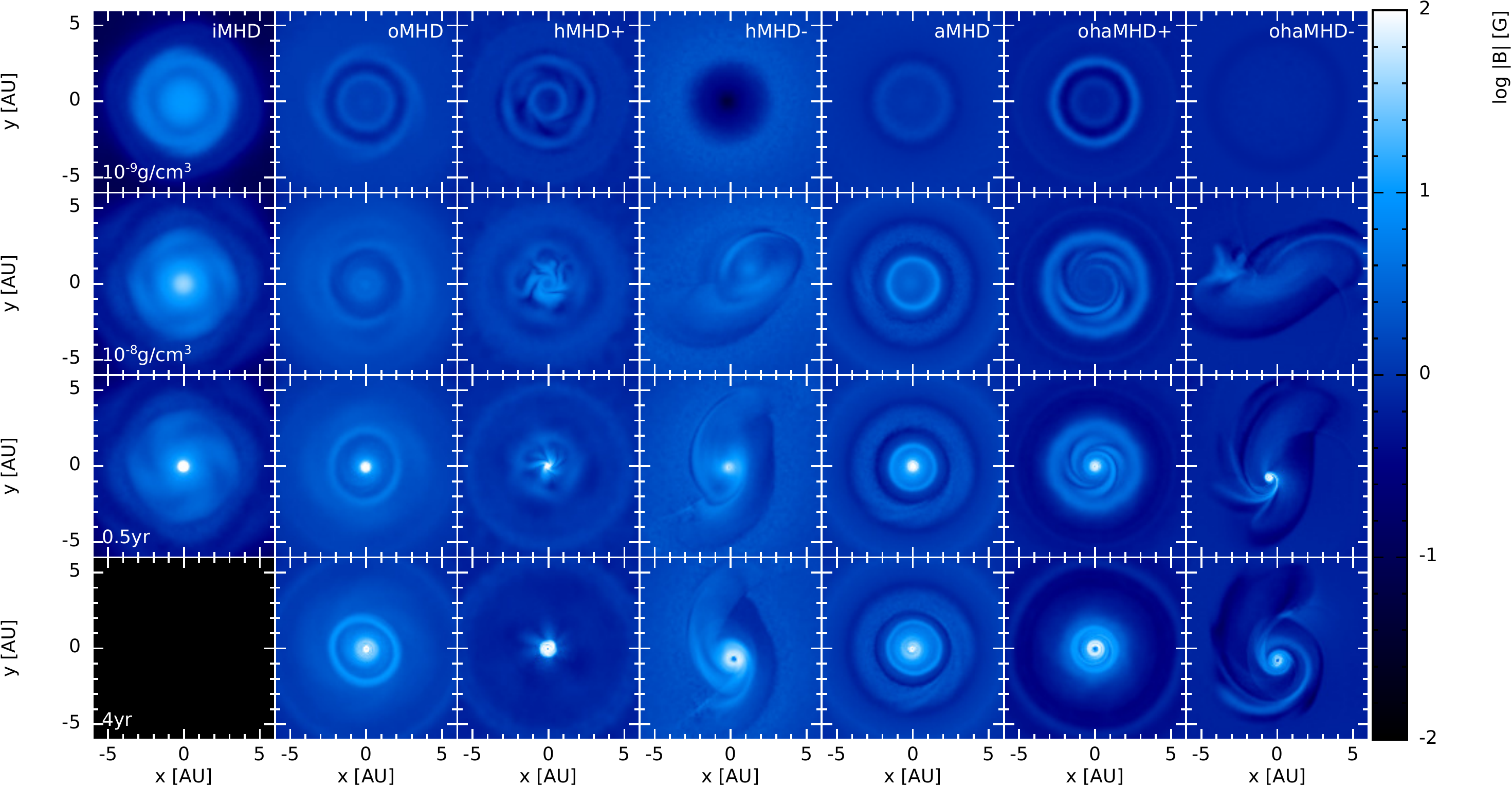}
\includegraphics[width=0.9\textwidth]{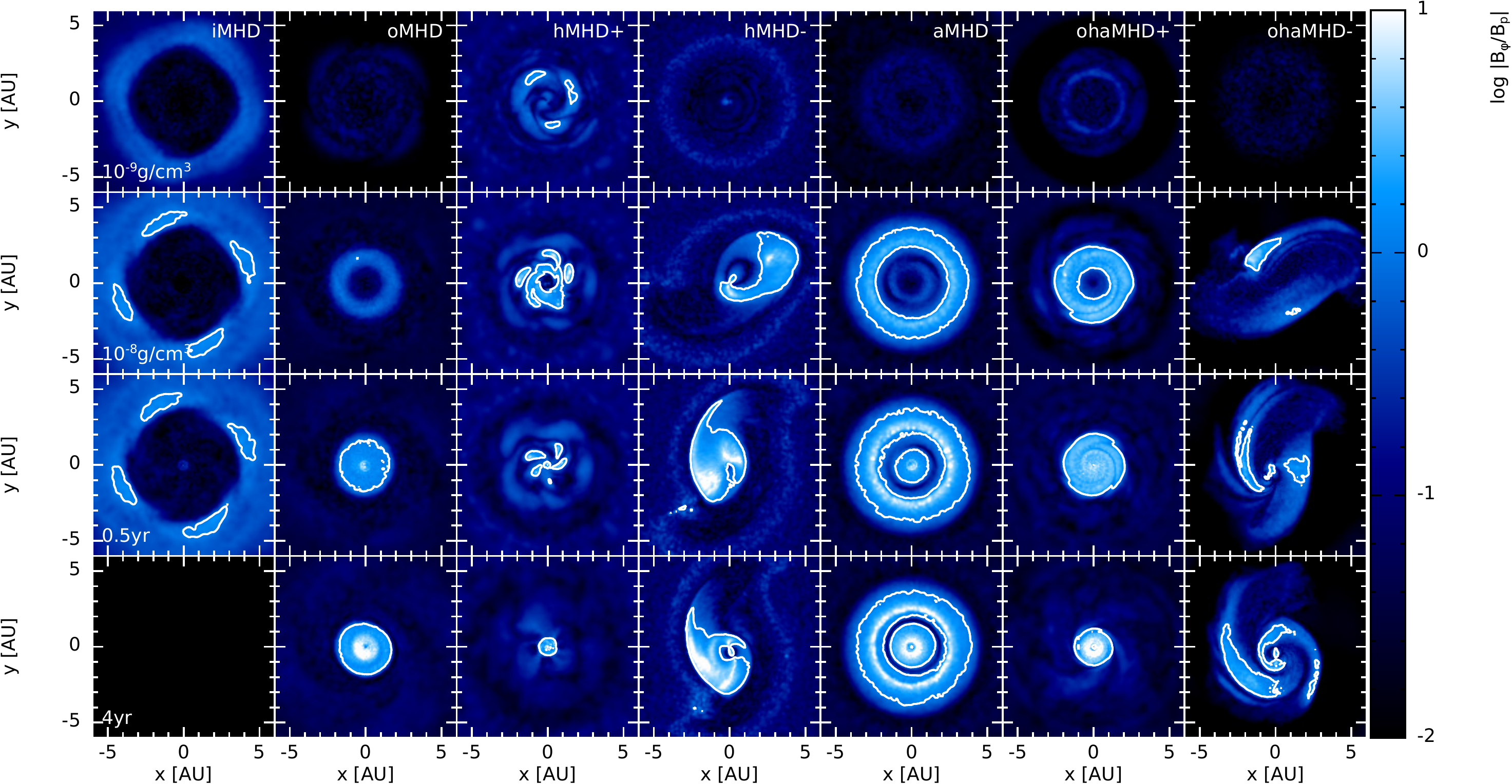}
\caption{Magnetic field strength (top) and ratio of toroidal-to-poloidal magnetic field strengths (bottom) slices through the cores perpendicular to the rotation axis at four epochs.  Model iMHD ended prior to \dtsc{4}, hence the blank frame.  Azimuthally symmetric magnetic walls form in oMHD and aMHD while a `pinwheel' structure forms in hMHD+.  When these processes are combined in ohaMHD+, a well-defined magnetic wall with a spiral interior forms.  The $m=2$ instabilities in hMHD- and ohaMHD- prevent the clear formation of a magnetic wall.  The well-defined magnetic walls (as in oMHD, aMHD and ohaMHD-) contain a strong poloidal component of the magnetic field, while the regions interior and/or exterior contain a strong toroidal component.}
\label{fig:B:xy}
\end{figure*}

The decoupling of the maximum and central field strengths occur in all models that include at least one non-ideal process, although a magnetic wall is only clearly visible in oMHD, aMHD and ohaMHD+, as shown in the top panel of \figref{fig:B:xy}.  An approximately azimuthally symmetric torus forms and persists in oMHD and aMHD, since there is no mechanism in these models to break the azimuthal symmetry.  The bottom panel of \figref{fig:B:xy} shows the ratio of the toroidal ($B_\phi = \left(xB_\text{y}-yB_\text{x}\right)/\sqrt{x^2+y^2}$) and poloidal ($B_\text{p} = \sqrt{B_\text{r}^2 + B_\text{z}^2}$) magnetic field strengths, which shows that the walls themselves are composed primarily of poloidal magnetic fields.  Exterior to the wall in aMHD at \rhoxge{-8} is a region dominated by the toroidal magnetic field.  For increasing radius within the first core, ambipolar diffusion become stronger as thermal ionisation becomes less important, which diffuses more of the magnetic field.  The weaker magnetic field yields a more rapid rotation of the core, which in turn twists the magnetic field from poloidal to toroidal, yielding a ratio of $\left| B_\phi/B_\text{p}\right| > 1$. After stellar core formation in oMHD and aMHD, a small disc forms interior to the magnetic wall, in which the toroidal magnetic field is the dominant component.

The Hall effect in hMHD- and ohaMHD- promotes rotation, resulting in azimuthal velocities higher than in aMHD.  However, these models undergo rotational $m=2$ instabilities \citepeg{Bate1998,Bate2011}, preventing them from forming well-defined magnetic walls during the first core phase or from generating a reasonable toroidal component of the magnetic field.  However, by \dtsc{4}, there is an azimuthally symmetric pileup of magnetic flux around the centre of the disc, where again, the maximum field strength is in this torus and not the centre.  

In model hMHD+, the vector evolution of the magnetic field results in a crude `pinwheel'  structure of the magnetic field by \rhoxeq{-8}, in which the magnetic field is mostly toroidal.  Although the magnetic flux piles up around the central region at later times and the toroidal field weakens, the pinwheel structure persists, resulting in a magnetic field structure that is not azimuthally symmetric, despite having an azimuthally symmetric density profile (recall \figref{fig:disc}).  In this model, the maximum and central field strengths are clearly decoupled, however, the magnetic wall is very near the centre.  

Finally, the magnetic field structure in ohaMHD+ is a result of all the non-ideal processes: an azimuthally symmetric torus that is dominated the poloidal field forms due to the diffusive terms, while the spiral structure (clearly visible at \rhoxge{-8}) is a result of the Hall effect.  The magnetic field in the spiral structure is primarily toroidal, and the toroidal-dominated region expands to encompass the entire region interior to the wall and the disc forms.  Therefore, all three non-ideal processes contribute to the complex magnetic field geometry during the star formation process. 

%--------------------------------------------------------------------------------
\section{Discussion}
\label{sec:disc}
%--------------------
\subsection{Magnetic field strength}
\label{sec:disc:mag}

For this study, we intentionally chose an initial mass-to-flux ratio of $\mu_0=5$.  This is to match our previous studies, and also as a compromise since modelling stronger magnetic fields is computationally more expensive due to shorter time-steps.

Our field strength agrees with that measured for the B335 Class 0 protostar by \citet{Maury+2018}; they also observe strong pinching of the magnetic field, in agreement with our results (see \figsref{fig:outflow:B}{fig:outflow:Bsc}).  However, most recent observations find star formation occurs in strongly magnetised regions with normalised mass-to-flux ratios of $0.5 \lesssim \mu \lesssim 3$, including 
the starless core L183 \citep{Karoly+2020}, %mu < 1
the Class 0 protostar Serpens SMM1 \citep{Hull+2017}, %mu sim 3
the massive star forming regions G240.31+0.07 \citep{Qiu+2014},  IRAS 4A \citep{GirartRaoMarrone2006} and L1157 \citep{Stephens+2013},  % mu \sim 1-2
and the 50 star forming regions studied by \citet{Koch+2014}; % mu sim 0.5 - 2
for reviews, see \citet{Crutcher1999}, \citet{HeilesCrutcher2005} and \citet{HullZhang2019}.  Determining the magnetic field strength is challenging, and cores tend to have larger mass-to-flux ratios than their envelopes \citep{Li+2014}, so care must taken when comparing the reported observed strength to those presented in numerical simulations.

Stronger magnetic fields increase the timescale for collapse and evolution, giving the physical processes (including the non-ideal processes) more time to impact the evolution.  In these simulations, the magnetic field is evolved as
\begin{eqnarray}
\label{eq:nimhd}
\frac{\text{d} \bm{B}}{\text{d} t} &=& \bm{\nabla} \times \left(\bm{v}\times\bm{B}\right) - \bm{\nabla} \times \left[  \eta_\text{OR}      \left(\bm{\nabla}\times\bm{B}\right)\right] \notag \\
                                                                                            &-& \bm{\nabla} \times \left[  \eta_\text{HE}       \left(\bm{\nabla}\times\bm{B}\right)\times\bm{\hat{B}}\right] \notag \\
                                                                                            &+&  \bm{\nabla} \times \left\{ \eta_\text{AD}\left[\left(\bm{\nabla}\times\bm{B}\right)\times\bm{\hat{B}}\right]\times\bm{\hat{B}}\right\},
\end{eqnarray}
where $\eta_\text{OR} \propto B^0$, $\eta_\text{HE} \propto B^1$ and $\eta_\text{AD} \propto B^2$ are the coefficients for Ohmic resistivity, the Hall effect and ambipolar diffusion, respectively.  Therefore, for stronger initial magnetic field strengths than modelled here, we expect the differences between the effects to be even more pronounced given their dependence on the magnetic field strength.  

%--------------------
\subsection{Comparison to the literature}
\label{sec:disc:lit}

This work follows from a long list of studies of isolated star formation from cloud collapse, as discussed in \secref{sec:intro}.  Most studies that investigate non-ideal MHD include Ohmic resistivity \textit{and} ambipolar diffusion and/or the Hall effect, although there are exceptions where only  the Hall effect \citepeg{KrasnopolskyLiShang2011} or only ambipolar diffusion \citepeg{Masson+2016} is included;  therefore, most studies investigate the effect of Ohmic resistivity plus another non-ideal process.  To the best of our knowledge, this paper and \citet{WursterPriceBate2016} are the only studies where the effects of Ohmic resistivity, ambipolar diffusion and the Hall effect are studied individually and compared directly to one another in a self-consistent framework.  Our current study complements its predecessors, as discussed below.  

\citet{BateTriccoPrice2014} investigated star formation using ideal MHD and several magnetic field strengths.  For decreasing magnetic field strengths, the first core outflows became broader while the stellar core outflows were qualitatively similar to one another with similar magnetic field strengths; i.e. first core outflows were dependent on the initial magnetic field strength while stellar core outflows were not.  Moreover, starting during the stellar collapse, the maximum magnetic field strength as a function of maximum density (similar to the top panel of our \figref{fig:BVrho}) showed very little difference between the models with \mueq{5} and 10, and the field strengths were only slightly higher than the model with \mueq{20}.  

\citet{Tomida+2013,TomidaOkuzumiMachida2015} model the gravitational collapse of a cloud using Ohmic resistivity and Ohmic resistivity + ambipolar diffusion (OR+AD).  We have general agreement with their results, although there are some notable differences; some of these differences may be a result of the initial conditions, such as the centrally condensed initial cloud and the seeded $m=2$ perturbation used by \citet{Tomida+2013,TomidaOkuzumiMachida2015} compared to our initial cloud of uniform density with a faster initial angular velocity.  Centrally condensing the initial sphere such that the central density is slightly lower than our uniform initial density increases the evolution time by nearly a factor of ten compared to our simulations, giving the non-ideal processes more time to affect the evolution.  Although the first core lifetimes are longer in \citet{Tomida+2013,TomidaOkuzumiMachida2015}, we agree that adding Ohmic resistivity and/or ambipolar diffusion increases the lifetime compared to ideal MHD.

By the formation of the stellar core, we find that oMHD and aMHD have formed small $r < 0.5$~au discs, and the disc in aMHD grows to $r \sim 3$~au by 4~yr after stellar core formation.  These sizes are in approximate agreement with \citet{Tomida+2013,TomidaOkuzumiMachida2015}, however, the disc in their OR+AD model forms during the first core phase, similar to the formation time of the discs in our hMHD- and ohaMHD-.  This early disc formation likely results in their conclusion that adding ambipolar diffusion drastically changes the structure of the first core.  Our models disagree and suggest that ambipolar diffusion alone \textit{does not} affect the structure of the first core.  Thus, ambipolar diffusion \textit{plus some other process/parameter} must be responsible for the change observed in  \citet{TomidaOkuzumiMachida2015}; 
it is possible that this change is a result of their longer first core lifetime resulting from an initially centrally condensed density profile (compared to our uniform density).  This degeneracy in processes/parameters and conflicting results amongst studies highlights the difficulty of understanding the precise effect and importance of each process/parameter.  

Finally,  \citet{Tomida+2013,TomidaOkuzumiMachida2015} concluded that the difference between the first core outflows in their models was a result of the lifetime of the first cores, whereas we have shown that the first core outflows are dependent on the included processes as well as first core lifetime (recall \secref{sec:outflow} and \figref{fig:outflow}).

The studies by \citet{Tsukamoto+2015oa,Tsukamoto+2015hall} primarily focused on disc formation.  They found small that $r \sim 1$~au discs formed with Ohmic resistivity or OR+AD, with the latter disc being slightly larger.  Our model aMHD yields a slightly larger disc than oMHD, thus the larger disc in the OR+AD model from \citet{Tsukamoto+2015oa} is likely primarily a result of ambipolar diffusion.  When the Hall effect is included in \citet{Tsukamoto+2015hall}, the disc is smaller ($r \lesssim 1$~au) if the magnetic field and rotation vectors are aligned, and the disc is $r \sim 20$~au when the vectors are anti-aligned.  Thus for all models, our disc sizes are in good agreement with theirs, despite the differences in initial conditions, including the initial angular velocity, initial magnetic field strength and microphysics describing non-ideal MHD.   Notably, we both initialised our clouds as 1~\Msun{} spheres of uniform density; therefore, based upon this work, \citet{Tomida+2013,TomidaOkuzumiMachida2015}, and \citet{Tsukamoto+2015oa,Tsukamoto+2015hall}, it appears that the initial density profile of the gas cloud plays an important role in determining its evolution.

%--------------------------------------------------------------------------------
\section{Summary and conclusion}
\label{sec:conc}

In this study, we investigated the effect of the individual non-ideal MHD terms on the formation of a protostar by following the gravitational collapse of a cloud core until stellar densities were reached.  All models were initialised as rotating, spherical 1~\Msun{} clouds of uniform density.  The models were threaded with a uniform magnetic field that was either aligned (+) or anti-aligned (-) with the rotation vector and had a normalised mass-to-flux ratio of 5.  We analysed seven models: an ideal MHD model (iMHD), models that include only Ohmic resistivity (oMHD), only ambipolar diffusion (aMHD), only the Hall effect (hMHD$\pm$), and all three non-ideal processes (ohaMHD$\pm$).  The models that included the Hall effect were modelled in duplicate (once with the magnetic field and rotation vectors aligned as in hMHD+ \& ohaMHD+, and once with them anti-aligned as in hMHD- \& ohaMHD-) due to the Hall effect's dependence on the magnetic field direction.  

Our main conclusions are as follows:

\begin{enumerate}
\item Including non-ideal MHD processes increased the lifespan of the first core phase compared to the ideal MHD model since these processes generally hindered the outward transport of angular momentum.
\item Including non-ideal MHD processes decreased the maximum magnetic field strength by at least an order of magnitude compared to the ideal MHD model; it also caused the maximum magnetic field strength to decouple from the central magnetic field strength (which is coincident with the maximum density) during the first core phase.  
\item Large, rotationally supported discs of $r \gtrsim 20$~au formed during the first core phase for hMHD- and ohaMHD-.  Small discs of $r \lesssim 4$~au formed in oMHD, aMHD and ohaMHD+ near the start of the stellar core phase.  No rotationally supported discs formed in iMHD or hMHD+; in the latter model, a counter-rotating pseudo-disc formed.  
Our results show that large discs only form when the Hall effect is included and the magnetic field and rotation vectors are initially anti-aligned, and that a bi-modality of disc sizes due to the Hall effect and the polarity of the magnetic field is expected, at least at early times.
\item First core outflows were launched in all models except hMHD- and ohaMHD-, suggesting that the Hall effect suppresses first core outflows when the magnetic field and rotation vectors are anti-aligned.  The launching of stellar core outflows was delayed in oMHD and aMHD compared to iMHD.  A very weak second core outflow was present in hMHD+.  Second core outflows were non-existent in hMHD- and ohaMHD$\pm$.  This suggests that the Hall effect suppresses stellar core outflows.
\item The Hall effect spun up/down the gas to modify the angular momentum budget near the core.  This caused the formation of a counter-rotating pseudo-disc in hMHD+ and large counter-rotating pockets in hMHD-.   The diffusive terms weakened the effect of the spin up/down such that no counter-rotating pockets formed in ohaMHD+ and only small, transient counter-rotating pockets formed in ohaMHD-.
\item Azimuthally symmetric magnetic tori (a.k.a. magnetic walls) formed in oMHD and aMHD during the first core phase.  The magnetic field structure in hMHD+ during this phase resembled a pinwheel.  These effects combined to form a well-defined torus with a spiral interior in ohaMHD+.  In all models (excluding iMHD), the non-ideal processes contributed to the magnetic flux piling up outside of the centre, leading to a decoupling of the central and maximum magnetic field strength.  
\end{enumerate}

Each individual non-ideal process affects the star formation process and its immediate environment.  While including only a single process may solve a single issue (e.g. ambipolar diffusion may prevent the magnetic braking catastrophe under selected initial conditions), it will lead to an incomplete picture of star formation.  These results reinforce our previous argument that ideal MHD yields an incomplete picture of star formation, but we now extend this to state that resistive MHD (i.e. including only Ohmic resistivity) also yields an incomplete picture since ambipolar diffusion and the Hall effect have a greater impact on the star formation process.

Although all three non-ideal MHD processes -- Ohmic resistivity, ambipolar diffusion and the Hall effect -- contribute to star formation, these results suggest that the Hall effect has the largest impact on the star formation process and the evolution of the surrounding environment given its ability to control the timescales, promote disc formation and suppress outflows.  

%--------------------------------------------------------------------------------
\section*{Acknowledgements}

We would like to thank the referee for useful comments that improved the quality of this manuscript.
JW and MRB acknowledge support from the European Research Council under the European Community's Seventh Framework Programme (FP7/2007- 2013 grant agreement no. 339248).  
JW and IAB acknowledge support from the University of St Andrews.
Calculations and analyses for this paper were performed on the University of Exeter Supercomputer, Isca, which is part of the University of Exeter High-Performance Computing (HPC) facility, and on the DiRAC Data Intensive service at Leicester, operated by the University of Leicester IT Services, which forms part of the STFC DiRAC HPC Facility (www.dirac.ac.uk). The equipment was funded by BEIS capital funding via STFC capital grants ST/K000373/1 and ST/R002363/1 and STFC DiRAC Operations grant ST/R001014/1. DiRAC is part of the National e-Infrastructure. 
%\texttt{These simulations were run on Isca and analysed on DiAL}.
Several figures were made using \textsc{splash} \citep{Price2007}.  

%----------------------------------------------------------------------------------------------------------------
\section*{Data availability}
The data for models iMHD and ohaMHD$\pm$ are openly available from the University of Exeter’s institutional repository at https://doi.org/10.24378/exe.607. 
The data for the remaining models will be available upon reasonable request.

\bibliography{nimhd.bib}
%--------------------------------------------------------------------------------
\label{lastpage}
\end{document}